\documentclass[twocolumn,secnumarabic,amssymb, nobibnotes, aps, prb,superscriptaddress]{revtex4-1}

\usepackage{amsthm,amssymb,amsmath,braket,mathdots}
\usepackage{graphicx}% Include figure files
\usepackage{dcolumn}% Align table columns on decimal point
\usepackage{bm}% bold math
\usepackage{hyperref}% add hypertext capabilities
\usepackage{natbib}
\usepackage{amsmath}
\usepackage{amsfonts}
\usepackage{color}
\usepackage[normalem]{ulem}
\DeclareMathOperator{\tr}{Tr}

\newcommand{\bit}{\begin{enumerate}}
\newcommand{\eit}{\end{enumerate}}
\newcommand{\zt}{\mathbb{Z}_2}
\newcommand*\diff{\mathop{}\!\mathrm{d}}
\newcommand{\non}{\nonumber}

\newcommand{\ve}[1]{{\boldsymbol{#1}}}

\newcommand{\kv}{\boldsymbol{k}}
\newcommand{\vel}{\boldsymbol{v}}
\newcommand{\vF}{v_{\text{F}}}

\newcommand{\mc}[1]{\mathcal{#1}}

\newcommand{\mac}{\mathcal}
\newcommand{\jcoeff}{\boldsymbol{\delta}_\epsilon}
\newcommand{\nue}{\nu_\epsilon}
\newcommand{\straintensor}{\underline{\boldsymbol{\mathcal{E}}}}

\newcommand{\Gb}{G_{\text{b}}}
\begin{document}

\preprint{APS/123-QED????????????}

\title{Current Switching of Valley Polarization in Twisted Bilayer Graphene}
% Current Switching of Valley Polarization in Twisted Bilayer Graphene Aligned with Hexagonal Boron Nitride
% Current Switching of Valley Polarization in moir\'e Heterostructure
\author{Xuzhe Ying}
\affiliation{Department of Physics and Astronomy, University of Waterloo, Waterloo, Ontario,  N2L 3G1, Canada}
\affiliation{School of Physics and Astronomy, University of Minnesota, Minneapolis, MN 55455, USA}
\affiliation{Kavli Institute for Theoretical Physics, University of California, Santa Barbara, CA 93106, USA}

\author{Mengxing Ye}
\affiliation{Kavli Institute for Theoretical Physics, University of California, Santa Barbara, CA 93106, USA}

\author{Leon Balents}
\affiliation{Kavli Institute for Theoretical Physics, University of California, Santa Barbara, CA 93106, USA}
	
\date{\today}% It is always \today, today,
% but any date may be explicitly specified
	
\begin{abstract}
  Twisted bilayer graphene (TBG) aligned with hexagonal boron nitride
  (h-BN) substrate can exhibit an anomalous Hall effect at 3/4 filling
  due to the spontaneous valley polarization in valley resolved
  moir\'e bands with opposite Chern
  number~\cite{serlin2020intrinsic,sharpe2019emergent}. It was
  observed that a small DC current is able to switch the valley
  polarization and reverse the sign of the Hall
  conductance~\cite{serlin2020intrinsic,sharpe2019emergent}. Here, we
  discuss the mechanism of the current switching of valley
  polarization near the transition temperature, where  bulk
  dissipative transport dominates. We show that for a sample with 
  rotational symmetry breaking, a DC current may generate an electron density
  difference between the two valleys (valley density difference). The
  current induced valley density difference in turn induces a first
  order transition in the valley polarization. We emphasize that the
  inter-valley scattering plays a central role since it is the channel
  for exchanging electrons between the two valleys. We further
  estimate the valley density difference in the TBG/h-BN system with a
  microscopic model, and find a significant enhancement of the
  effect in the magic angle regime.
	%Our result shows that due to the small Fermi velocity and the huge Moire unit cell, a small current is enough to switch the valley polarization.

\end{abstract}

\maketitle

        %%%%%%%%%%%%%%%%%%%%%%%%%%%%%%%%%%%%%%%%%%%%%%%%%% 
\section{Introduction}

Spontaneous ferromagnetism in a purely itinerant electron gas without
local moments is an old theoretical idea first imagined by Stoner in
the 1930s\cite{stoner1938collective}.  Realizations of this ideal have
not been easy to find.  The clearest and best studied example is
probably quantum Hall
ferromagnetism\cite{sondhi1993skyrmions,macdonald1996skyrmions}, where
the Stoner instability is enable by the flatness of Landau levels
induced by an orbital magnetic field.  Quantum Hall ferromagnetism is,
however, not ultimately true ferromagnetism insofar as time-reversal
symmetry is from the outset strongly and explicitly broken by a large
applied magnetic field.  Recently, purely itinerant ferromagnetism has
been observed in zero magnetic field in twisted bilayer graphene
(TBG), adding to the host of exotic phenomena in this system when
twisted near the ``magic angle'' at which the moir\'e bands becomes
exceptionally
flat\cite{sharpe2019emergent,serlin2020intrinsic,cao2018correlated,cao2018unconventional,PhysRevB.98.075154,PhysRevB.98.085435,PhysRevB.99.075127,balents2020superconductivity,PhysRevLett.121.126402,Zhang2019hBN,PhysRevX.8.041041,nandkishore2012chiral,PhysRevB.101.224513,PhysRevB.102.125120,PhysRevLett.121.087001,PhysRevX.8.031089,yankowitz2019tuning,lu2019superconductors,cao2020nematicity,jiang2019charge,PhysRevB.99.035111,PhysRevB.100.035115,saito2020independent,xie2019spectroscopic,tschirhart2020imaging,PhysRevLett.122.246401,polshyn2020electrical}.
The most dramatic signatures of itinerant ferromagnetism occur in TBG
samples aligned to an hexagonal-Boron Nitride (h-BN) substrate at
$3/4$ filling\cite{serlin2020intrinsic,sharpe2019emergent}.  Here the
ferromagnetism observed below the Curie temperature of $5-8K$ is
observed via an anomalous Hall effect (AHE) -- a zero field hysteretic Hall
resistivity -- that evolves into a quantized value of
$\rho_{xy}=h/e^2$ at low temperature: a quantum anomalous Hall effect
(QAHE).  The existence of the QAHE, which has been discussed extensively theoretically\cite{PhysRevB.99.075127,PhysRevLett.124.166601,ochi2018possible}, is related to two aspects of TBG.  First,
graphene itself has incipient valley Chern number associated to its
Dirac points, which is created even
in a single layer by an infinitesimal perturbation breaking inversion
or $C_2$ rotation or time-reversal $\mac{T}$ symmetries.  In TBG this
extends to bands formed from both layers, and with h-BN to break the
$C_2 \mac{T}$, and the 4 active moir\'e bands acquire unit Chern
number with sign that is  opposite for conduction and valence bands
and opposite for each valley.  The second necessary aspect for (Q)AHE
in TBG is symmetry breaking.  An AHE then is observed when the
difference of occupation of the two valleys  -- the valley
polarization $\Phi_{\text{v}}$ -- becomes non-zero. This signifies spontaneous breaking
of $\mac{T}$ symmetry and defines ferromagnetism.  The presence of
QAHE implies that {\em at low temperature}  both spin and valley symmetries
are broken,  and both are fully polarized.    Note that for the AHE at
temperatures close to the Curie point, the valley polarization $\Phi_{\text{v}}$ is
the order parameter, and spin symmetry breaking is not essential.    The sign of the Hall conductivity is
determined by the valley polarization, so that tuning the latter
controls the former.

Interestingly, in experiment, the sign of the Hall conductance can be controlled by
either an external magnetic field or a bias electric
field/current\cite{sharpe2019emergent,serlin2020intrinsic}.   Similar
hysteresis curves were observed on 
sweeping either the magnetic field or  the DC current, indicating an
apparent first order transition in the valley polarization, similar to
the way in which the magnetic field affects the magnetization in the
Ising model.

The sensitive magnetic field control of the valley polarization and
thus the Hall conductance has been well explained by linear free
energy dependence associated to the giant orbital magnetic moment of
the moir\'e Bloch
electrons\cite{serlin2020intrinsic,sharpe2019emergent,chang2008berry,RevModPhys.82.1959},
which is closely related to the large Berry curvature of the valley
Chern bands.  The mechanism for current switching of the Hall
conductance remains less clear.  Several proposals have been made for
this mechanism at low temperature\cite{he2020giant,serlin2020intrinsic,MacDoanld2020}.  Here, we focus on higher
temperatures near but below the Curie point.  In experiment, the sign
of the Hall conductance remains sensitive to the DC current in this
regime, where indeed experiments are significantly easier and more
reproducible, due to weakness of hysteresis.  At these temperatures,
the bulk is dissipative $\sigma_{xx}\neq 0$ and $\sigma_{xy}$ is not
quantized, and indeed the Hall angle
$\theta_H \sim \sigma_{xy}/\sigma_{xx} \ll 1$ is small.

In this highly conducting situation, it may be tempting to
make analogies to current switching of common metallic ferromagnets,
where it is usually ascribed to ``spin torque''.  However, some
important differences are evident.  First, in TBG, the magnetization
itself is primarily orbital, and indeed we do not expect significant
spin polarization near the Curie point.  Second, a related point is
that  normal ferromagnets have an approximate spin-rotation symmetry
(arising from weak spin-orbit coupling)
and the ferromagnetism is described by a vector order parameter with
weak anisotropies; in contrast, in TBG the valley polarization is
Ising-like and not a vector.  The Ising symmetry that changes the sign
of the valley polarization is just $\mac{T}$.  Finally, in clean TBG there
is to an excellent approximation a valley conservation symmetry.  This
is not the symmetry spontaneously broken by the AHE, but rather it implies that the
valley polarization order parameter is approximately {\em conserved}.

In this article, we report a mechanism that takes these features into
account and leads to the control of valley polarization by a DC
current. As a consequence of the quasi-conservation of the order
parameter, in this mechanism, inter-valley scattering plays a central
role.  We first study the dynamics of the valley polarization order
parameter (VPOP) near the Curie temperature $T_c$ by obtaining its
equation of motion (EoM). The EoM shows that any mechanism which can
generate an electron density difference between the two valleys in the
non-interacting model can induce a first order transition of the VPOP.
By solving the semiclassical Boltzmann equation, we show that the
valley density difference can be generated by a DC current with
inter-valley scattering that breaks the rotational symmetry to
$\mc{C}_{1z}$.  We find that the valley density difference is
proportional to the current density, the inverse of the Fermi velocity
and the strength of the rotational symmetry breaking. We make a
specific estimate for the magnitude of the effect for TBG aligned with
h-BN (denoted as TBG/h-BN system hereafter), and demonstrate two
sources of enhancement in comparison with the single layer graphene. First,
 we show an enhancement of the
effective strain from $\epsilon$ in single layer graphene to
$\epsilon/\theta_w$ in TBG with twist angle $\theta_w$. Second, near
the magic twist angle, the Fermi velocity is significantly reduced
from $10^6$ m/s to around $10^4-10^5$ m/s. Combining the two effects,
the enhancement of the current induced valley density difference is on
the order of $10^3$. Thus, the valley polarization is very sensitive
to the applied DC current.
	
The rest of the paper is organized as follows. In
Sec.~\ref{Sec:ValleyPolarization}, we introduce the model, discuss the
dynamics and steady state solution of the valley polarization order
parameter, and demonstrate how the it can be controlled by a DC
current qualitatively. To obtain this relation quantitatively, in
Sec.~\ref{Sec:CurrentInducedValleyDensityDifference}, we present the
Boltzmann equation and estimate the inter-valley scattering rate for
the TBG/h-BN system. The technical details are postponed to the
Appendix. App.~\ref{Sec:VPOPDynamicsFromKeldysh} derives the dynamics
of VPOP within the Keldysh formalism. App.~\ref{Sec:ModelingofTBG}
presents the details of the modeling of the TBG/h-BN system.
% 	The rest of the article is structured as: In Sec.~\ref{Sec:ValleyPolarization}, we introduce the general framework and the dynamics of the valley polarization and discuss qualitatively how the valley polarization can be controlled by a DC current; Sec.~\ref{Sec:CurrentInducedValleyDensityDifference}, we present the Boltzmann equation and analyze the inter-valley scattering to study the valley density difference under a bias DC current (or electric field) and discuss the limiting factors. The technical details are postponed to the Appendix: App.~\ref{Sec:VPOPDynamicsFromKeldysh} derives the dynamics of VPOP within Keldysh formalism; App.~\ref{Sec:ModelingofTBG} presents the details of the modeling for TBG.
	%%%%%%%%%%%%%%%%%%%%%%%%%%%%%%%%%%%%%%%%%%%%%%%%%%
\section{Valley Polarization}
\label{Sec:ValleyPolarization}
\begin{figure}[tb]
  \centering
  \includegraphics[width=\linewidth]{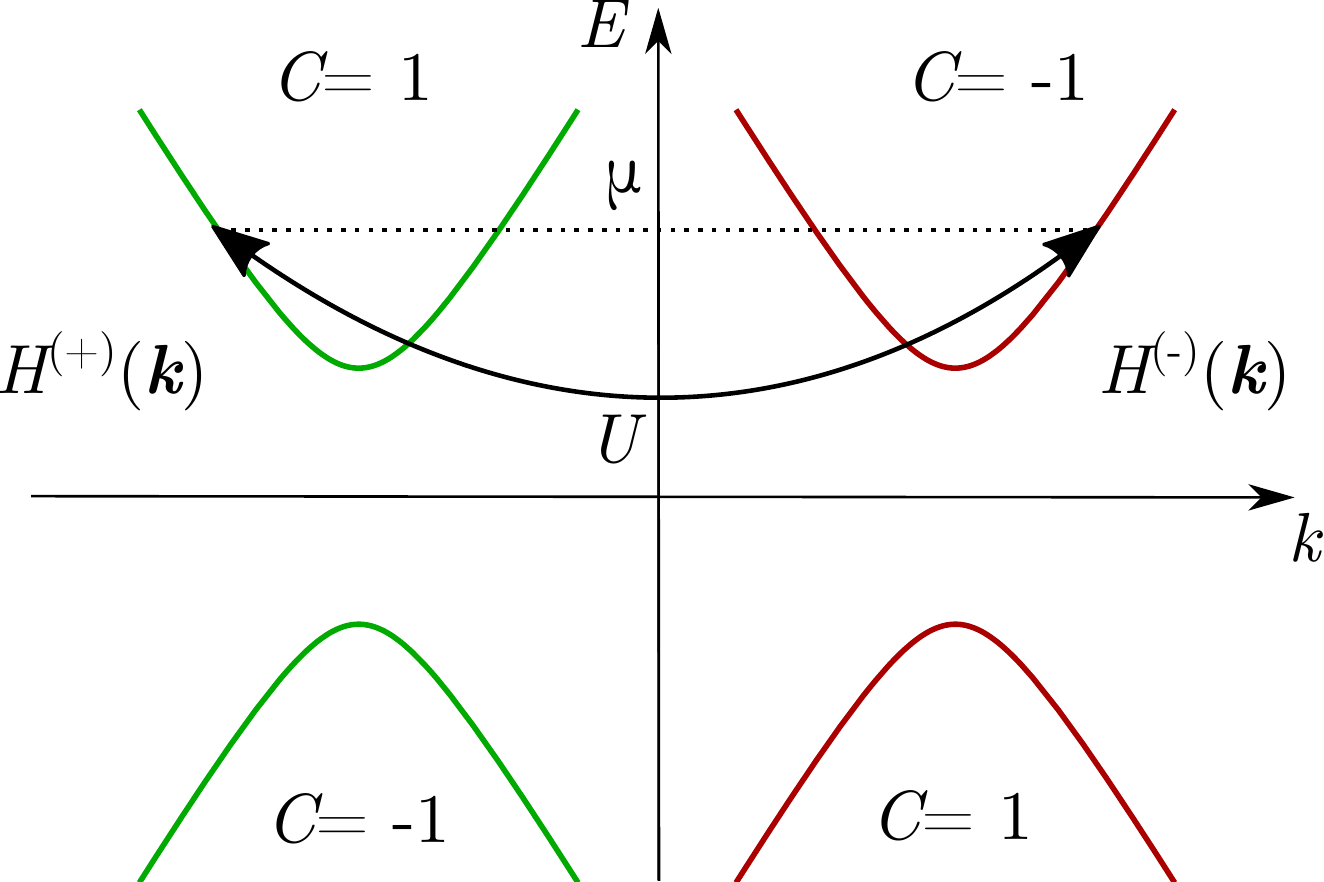}
  \caption{Schematic of model with two valleys, with a Stoner-type interaction. The two valleys are time reversal related, so that they carry opposite Chern numbers.}
  \label{Fig:TwoValleyModel}
\end{figure}

In this section, we discuss the dynamics of the valley polarization order parameter. We employ the nonequilibrium Keldysh approach\cite{kamenev2011field}, and obtain the equation of motion (EoM) for the valley polarization order parameter (VPOP) due to the interplay between the interaction and the external current.   This construction demonstrates the relation between the current (or magnetic field) induced polarization of non-interacting electrons, $\Delta n_0$, and the true polarization (the VPOP) $\Phi_{\text{v}}$, including interactions.

%%%%%%%%%%%%%%%%%%%%%%%%%%%%%%%%%%%%%%%%%%%%%%%%% 
\subsection{The Model}

In this article, we consider a model with 2 copies of Chern insulators labeled as $s=\pm$, Fig.~\ref{Fig:TwoValleyModel}, with the following free fermion Hamiltonian:
\begin{equation}
  H_0=\sum_{i=\pm}\psi^{(s)\dagger}_{\boldsymbol{k}}H^{(s)}(\boldsymbol{k})\psi^{(s)}_{\boldsymbol{k}}.
\end{equation}
The two copies of Chern insulators are further assumed to be related
by the time reversal symmetry $\mathcal{T}$, such that
$\mathcal{T}H^{(\pm)}(\boldsymbol{k})\mathcal{T}^{-1}=H^{(\mp)}(-\boldsymbol{k})$. With
the restriction from the time reversal symmetry, the two conduction
bands (as well as the valence bands) carry opposite Chern number,
Fig.~\ref{Fig:TwoValleyModel}. This model may be considered as a low
energy effective model for the TBG/h-BN system~\cite{Zhang2019hBN},
when only the 2 active moir\'e bands (in sublattice space) in each
valley is included, and each copy of a Chern insulator corresponds to a valley. Thus, the two copies of Chern insulators are referred to as two valleys in the rest of the article.
% This model can well describe the free electronic states in TBG, in the sense that each copy of Chern insulator corresponds a valley. Thus, the two copies of Chern insulators are referred to as two valleys in the rest of the article.

To model the interaction induced valley polarization, we restrict the interaction to the inter-valley density-density channel only:
\begin{equation}
  H_{\text{int}}= U\,\int \diff \ve{x}\,n^{(+)}(\ve{x})n^{(-)}(\ve{x}),
\end{equation}
where $U$ is the interaction strength that we approximate as a
constant, and $n^{(\pm)}$ is the electron density of the $\pm$
valleys. This is a caricature of the inter-valley component of the
Coulomb interaction.  We expect that the precise form of the
interaction is not important, so long as the symmetries of the problem
(time-reversal and valley conservation) are respected, as we will be
primarily interested in low energy quantities in the vicinity of the
Curie point.  At strong interaction $U>U_c$, the valley polarization
develops spontaneously at low temperature. The critical interaction
$U_c$ can be estimated to be the inverse of the density of states at
Fermi level according to the Stoner criteria, i.e.\
$U_c\sim\nu^{-1}$\cite{altland2010condensed}.
% MY: The electron band is not filled, just valley polarization develops. 
% When the interaction is strong enough $U>U_c$, the electrons would fill one of the two valleys to minimize the interaction energy. The critical interaction is given by the inverse of the density of states at Fermi level, $U_c=\nu^{-1}$. There is also a critical temperature $T_c$, below which spontaneous valley polarization is possible.

Note that the spin degrees of freedom are ignored in our study.  As
discussed in the Introduction, the AHE requires only valley and not
spin polarization.  Furthermore, in the vicinity of the Curie point,
there is unlikely to be substantial spin polarization, since with SU(2)
spin symmetry the Mermin-Wagner theorem\cite{mermin1967absence}
prohibits any $T>0$ order, and SU(2) spin symmetry is broken extremely
weakly by tiny spin-orbit and dipolar effects.

%%%%%%%%%%%%%%%%%%%%%%%%%%%%%%%%%%%%%%%%%%%%%%%%% 
\subsection{Steady State Solution of the Valley Polarization Order Parameter}

We now obtain the EoM of the VPOP using the non-equilibrium Keldysh
approach.  Details of the derivations are given in
App.~\ref{Sec:VPOPDynamicsFromKeldysh}. It is essential to introduce a proper scattering mechanism in order to establish a steady state subject to an electric field. We consider short ranged disorder described by an impurity potential $V^{\text{imp}}(\boldsymbol{x})$, which induces both intra- and inter-valley scattering [see Eq.~\eqref{appeq:disorder}].

 Near the transition temperature ($T\sim T_c$), the EoM can be
 expressed as an expansion in powers of the VPOP $\Phi_{\text{v}}$.  It takes the form
\begin{equation}
  \alpha_2(\omega,q)\Phi_{\text{v}}+\alpha_4\Phi_{\text{v}}^3+\Delta n_0=0,
  \label{Eq:EoMOrderParameter}
\end{equation}
which should be regarded as somewhat symbolic, with the time and space
dependence expressed in the first term in Fourier space, while the
second and third terms may be considered approximately local.  To the
leading order in $\vert T-T_c\rvert$ and external bias electric field, quadratic
terms $\sim \alpha_3h^2$ can be ignored (they vanish in equilibrium
without any symmetry breaking field). In the static limit for the
homogeneous order parameter, i.e.\ $\omega= 0$ and then $\ve{q}\rightarrow 0$, this
reduces to the standard expression that mimics the 1st order Ising phase transition in an external field, i.e.\
\begin{align}
  (r-r_c)\Phi_{\text{v}} -\alpha_4 \Phi_{\text{v}}^3=\Delta n_0,
  \label{eq:VPOPsol}
\end{align}
where $\Delta n_0$ is the valley density difference that would be
induced by the bias electric field in the absence of interactions (and
hence is smooth near $T_c$ because the transition is induced by
interactions).  The quantity
$(r-r_c)=-\alpha_2(0,0)\sim (T/T_c-1)$~\cite{altland2010condensed}
changes sign across the equilibrium transition. The cubic coefficient
$-\alpha_4\sim |\nu''(\epsilon_F)|U^3$ is positive definite,
corresponding to a bounded equilibrium free energy, and ensures the
stability of the state across the transition.
% So we have %\cite{Fernandes2012}
%	\begin{align}
          %           \alpha_4 h^3+(r-r_c) h+\Delta n_0=0.
          %           \label{eq:VPOPsol}
          %         \end{align}
By construction, the VPOP describes the expectation value of the
valley density difference,
$\Phi_{\text{v}}= \left(\langle n^{(+)}\rangle-\langle n^{(-)}\rangle\right)$ (see
App.~\ref{Sec:VPOPDynamicsFromKeldysh}).  Keep in mind that
$\Delta n_0$ is valley density difference induced by the external bias
field alone without interactions, while the VPOP $\Phi_{\text{v}}$ describes the
valley density difference with both the external bias field and
interactions taken into account.

          %           The solution for the VPOP should then satisfy
          %           \begin{align}
          %           \alpha_4 h^3+(r-r_c) h-\alpha_0 \Delta n_0=0.
          %           \label{eq:VPOPsol}
          %         \end{align}
          %         Eq.~\eqref{eq:VPOPsol} describes the steady state solution of the VPOP in an external electric field near $T_c$,	where the dimensionless parameters are $\alpha_4 \sim \nu''(\epsilon_F)U^3$, $\alpha_2=(r-r_c)\sim(1-T/T_c)$\cite{altland2010condensed}. To the leading order in $(r-r_c)$ and external bias electric field, the quadratic terms $\sim \alpha_3h^2$ has been ignored.

\begin{figure}[tb]
	\centering
	\includegraphics[width=\linewidth]{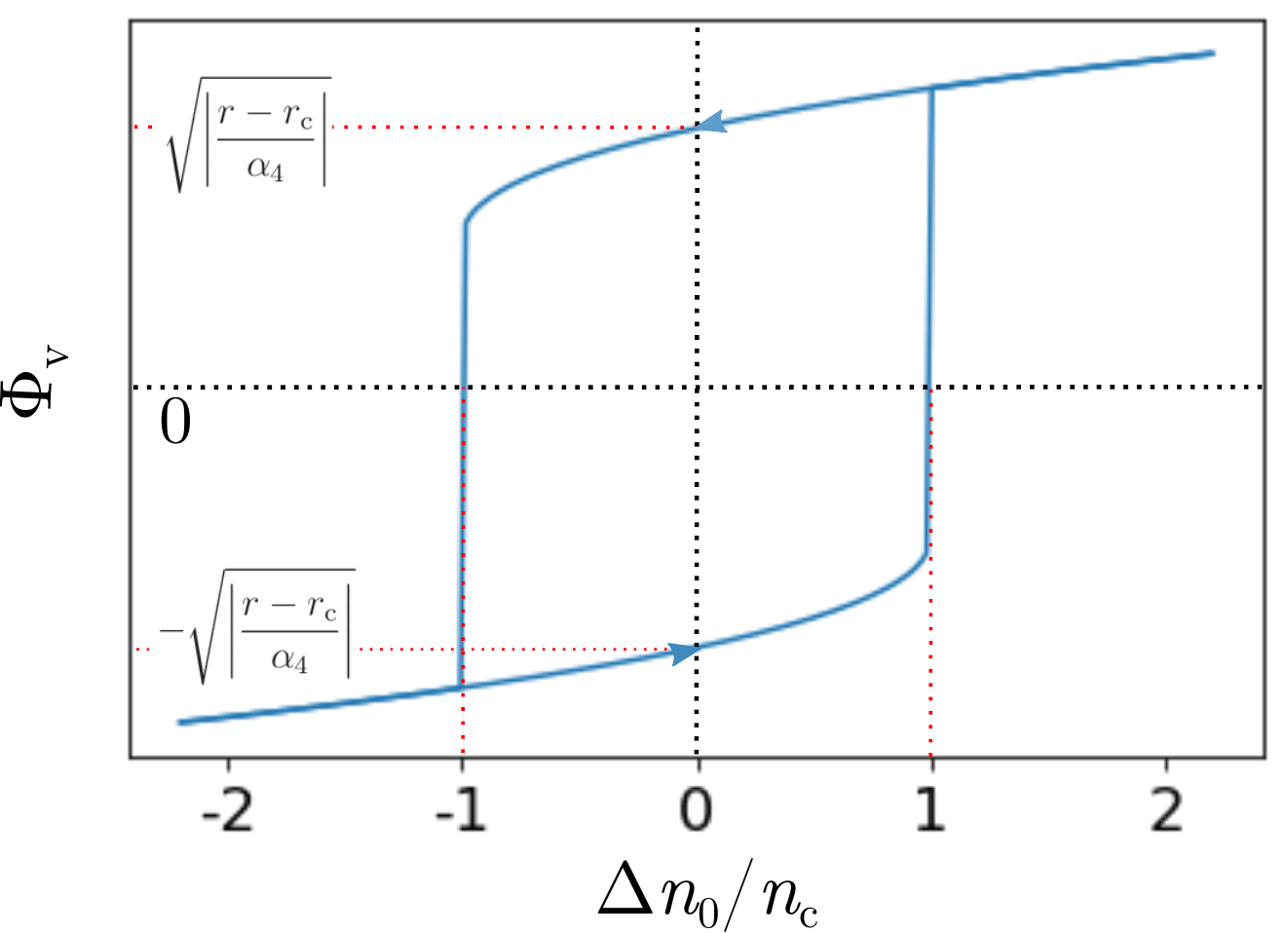}
	\caption{Hysteresis curve for valley polarization $\Phi_{\text{v}}$ upon tuning the valley density difference $\Delta n_0$ at $T<T_c$. The coercive valley density difference is $n_c=-2\alpha_4\left(\frac{r-r_c}{3\alpha_4}\right)^{3/2}$.}
	\label{Fig:Hysteresis}
\end{figure}

Without the bias electric field, $\Delta n_0=0$, Eq.~\eqref{eq:VPOPsol} describes  spontaneous $\zt$ symmetry breaking in equilibrium when $T<T_c$ [$(r-r_c)<0$], with $\Phi_{\text{v}}=\pm\sqrt{|\frac{(r-r_c)}{\alpha_4}|}$. Non-zero $\Delta n_0$ explicitly breaks the $\zt$ symmetry, and selects the $+$ or $-$ VPOP, when $\Delta n_0$ is positive or negative, respectively. By tuning $\Delta n_0$, one recovers the hysteresis curve, Fig.~\ref{Fig:Hysteresis}. The coercive valley density difference is given by $n_c=-2\alpha_4\left(\frac{r-r_c}{3\alpha_4}\right)^{3/2}$.

To address how the external bias electric field controls the valley
polarization, we discuss below how $\Delta n_0$ depends upon the bias
electric field $\ve{E}$, or equivalently the current density
$\ve{j}$. Importantly, breaking lattice rotational symmetry is necessary
to generate any valley density difference by the current $\ve{j}$.
% generated by the external bias is only possible if the rotational symmetry is broken. 
This is because the bias electric field and current, $\boldsymbol{E}$ and $\boldsymbol{j}$, are vectors in 2D. To make a non-zero scalar, $\Delta n_0$, another vector is needed.   This means that there is a particular direction in the sample. Thus, the (discrete) rotational symmetry has to be broken.

Moreover, by dimensional analysis, one can easily show that the valley density difference generated by an applied DC current should be given by:
\begin{align}
  \Delta n_0 \simeq \frac{a}{ev_{\text{F}}}j_x+\frac{b}{ev_{\text{F}}}j_y = \frac{1}{e v_\text{F}} \ve{j}\cdot \boldsymbol{\delta}_\epsilon
  \label{Eq:ValleyDensityDifferenceGeneral},
\end{align}
which is proportianl to the current density in 2D, $\boldsymbol{j}=(j_x,j_y)$, and inverse of the Fermi velocity $v_{\text{F}}$. The dimensionless parameters, $a(b)$ or $\boldsymbol{\delta}_\epsilon$, are related to the broken rotational symmetry. They are also highly dependent on the microscopic details of the system, which we do not attempt to address in depth in this article.

\begin{figure}[tb]
  \centering
  \includegraphics[width=0.9\linewidth]{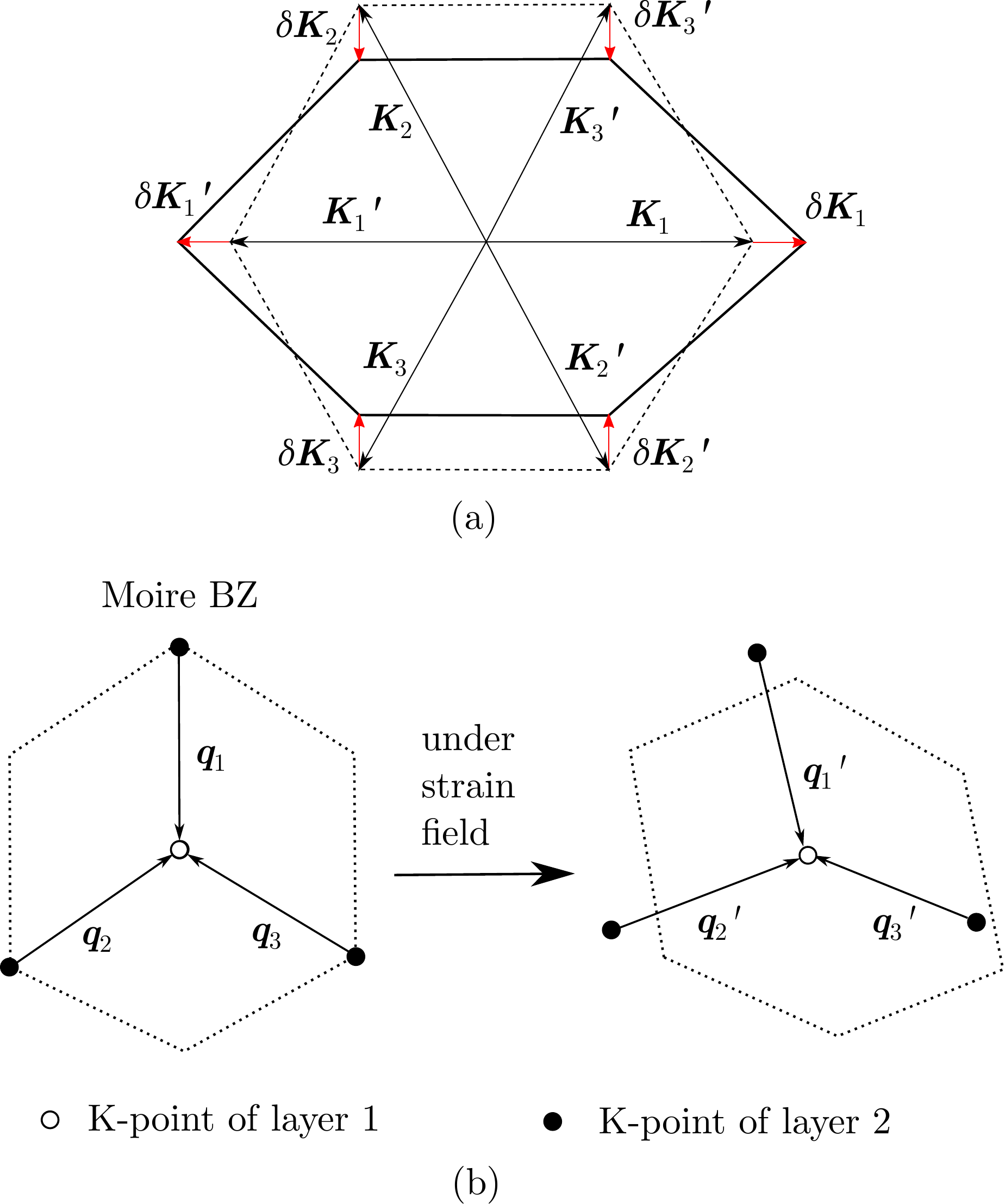}
  \caption{(a) Brillouin zone (BZ) of single layer graphene with uniaxial strain; (b) moir\'e Brillouin zone without (left) and with (right) uniaxial strain. The $\mathcal{C}_3$ symmetry of the unstrained moir\'e BZ is explicitly broken by the strain field.}
  \label{Fig:StrainAndBZ}
\end{figure}

\begin{figure}[tb]
	\centering
	\includegraphics[width=0.9\linewidth]{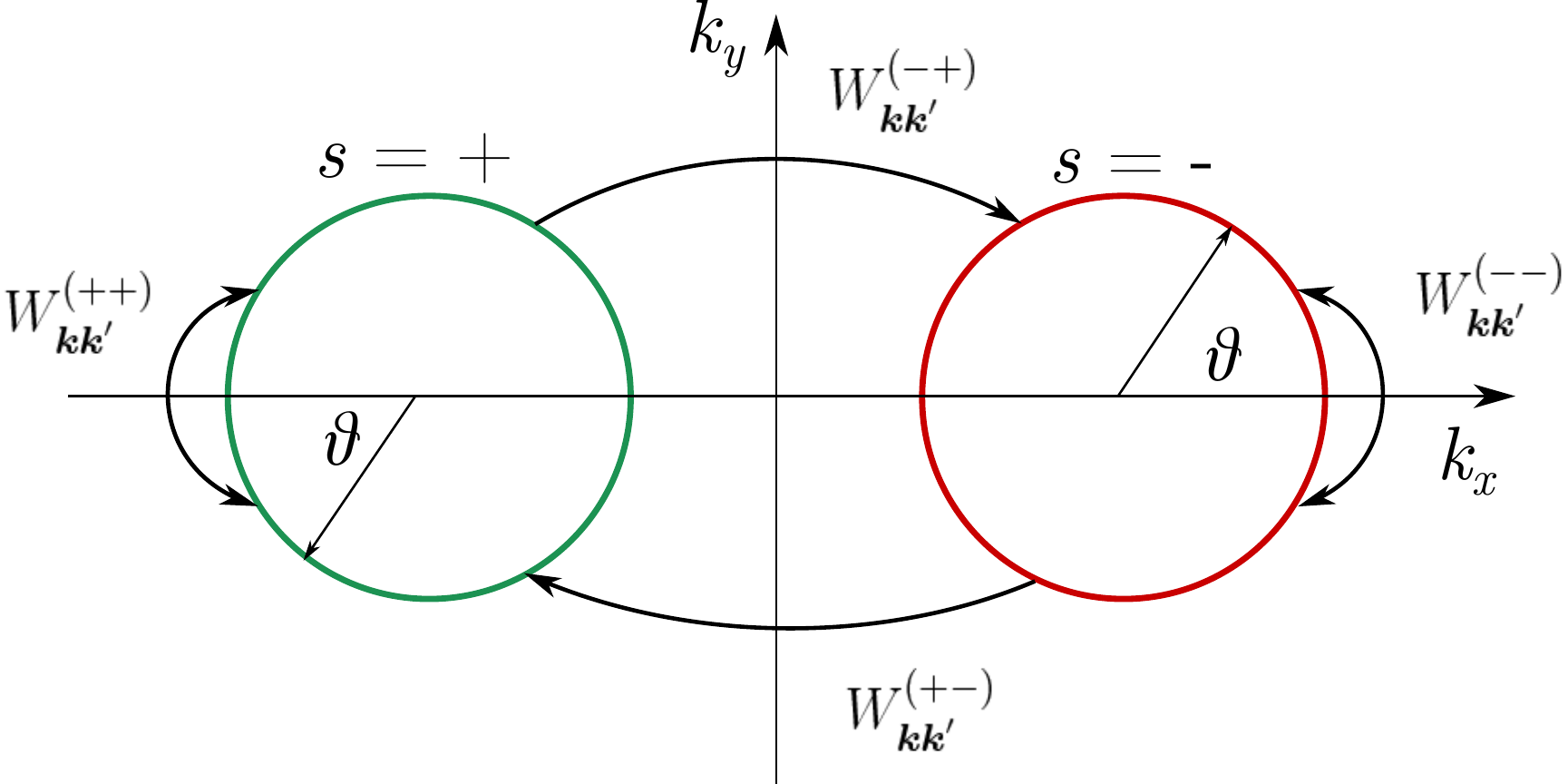}
	\caption{Impurity scattering between two Fermi pockets.}
	\label{Fig:SimplifiedScattering}
\end{figure}

Following the general discussion above, one may qualitatively argue
that in the TBG/h-NB system, the valley density difference generated
by a DC current can be quite significant for two reasons. First, the
small Fermi velocity of the flat bands near the magic angle increases
$\Delta n_0$ by a factor on the order of $10^2-10^3$.  Second, the
smallness of the moir\'e Brillouin zone enhances the proportional
effect of strain, as follows.  Strain results in anisotropy in the
electronic spectrum, reflected in a shift of the location of the Dirac
point, $\delta \boldsymbol{K}$, directional dependence of the Dirac
velocity, $\delta v_{\text{D}}$, etc. For single layer graphene, the anisotropy
can be characterized by a small parameter, for example the strain
strength,
$\epsilon\sim\frac{\left|\delta
    \boldsymbol{K}\right|}{\left|\boldsymbol{K}\right|},\ \frac{\delta
  v_{\text{D}}}{v_{\text{D}}}$, see Fig.~\ref{Fig:StrainAndBZ}(a).  For strained TBG, as in
Fig.~\ref{Fig:StrainAndBZ}(b), the shift of the Dirac points should be
compared with the size of the moir\'e BZ. Thus, the broken
$\mathcal{C}_3$ symmetry is actually characterized by
$\frac{\epsilon}{\theta_w}\sim\frac{\left|\delta\boldsymbol{K}\right|}{q}$,
where $\theta_w$, $q=\left|\boldsymbol{K}\right|\theta_w$ are the
small twist angle and distance between the adjacent Dirac points of
the two graphene layers due to the twist, an enhancement of a factor
of $1/\theta_w$ due to moir\'e physics.  These two effects enhance the
dimensionless parameters $\boldsymbol{\delta}_\epsilon$ in
Eq.~(\ref{Eq:ValleyDensityDifferenceGeneral}), which are thus not
necessarily small, and the effect may be quite significant.

\section{Current induced valley density difference}
\label{Sec:CurrentInducedValleyDensityDifference}
In this section, we employ the semi-classical Boltzmann equation to demonstrate how a DC current may induce a valley density difference $\Delta n_0$ for models without rotational symmetry, and estimate the dimensionless coefficient $\boldsymbol{\delta}_\epsilon$ for TBG aligned with h-BN.

\subsection{Toy Model and the Semi-classical Formalism}
\label{subsec:SBE}

In this subsection, we demonstrate the effect of inter-valley scattering on the current induced valley density difference $\Delta n_0$ by solving the semi-classical Boltzmann equation (SBE). We simplify the Fermi surface at each valley as a circular Fermi pocket as shown in Fig.~\ref{Fig:SimplifiedScattering}, and will argue later that this simplification doesn't change the result qualitatively.  The calculation is carried out in the absence of interactions, i.e. in the paramagnetic phase, so that  time reversal symmetry is present, which imposes $\epsilon^{s}(\kv)=\epsilon^{\bar{s}}(-\kv)$ and $\vel^{s}(\kv)=-\vel^{\bar{s}}(-\kv)$, where $s=\pm$ is the valley index with $s\neq \bar{s}$. No other point group symmetries are assumed.  

The SBE within the presence of a bias electric field $\boldsymbol{E}$ is given by\cite{kamenev2011field,LandauKinetics}:
\begin{widetext}
  \begin{equation}
    \partial_tf^{(s)}_{\boldsymbol{k}}+\boldsymbol{v}^{(s)}_{\boldsymbol{k}}\cdot\partial_{\boldsymbol{x}}f^{(s)}_{\boldsymbol{k}}+e\boldsymbol{E}\cdot\partial_{\boldsymbol{k}} f^{(s)}_{\boldsymbol{k}}=\sum_{s^{\prime}=\pm}\int \diff\Gamma^{\prime}W^{(ss^{\prime})}_{\boldsymbol{k}\boldsymbol{k}^{\prime}}\left(f^{(s^{\prime})}_{\boldsymbol{k}^{\prime}}-f^{(s)}_{\boldsymbol{k}}\right)\delta(\epsilon^{(s^{\prime})}_{\boldsymbol{k}^{\prime}}-\epsilon^{(s)}_{\boldsymbol{k}})\ =I_{\text{intra}}^{(s)}[f_{\boldsymbol{k}}]+I_{\text{inter}}^{(s)}[f_{\boldsymbol{k}}].
    \label{Eq:SBE0}
  \end{equation}
\end{widetext}
The measure in the collision integral is defined as $\diff\Gamma^{\prime}=\frac{d^2k^{\prime}}{(2\pi)^2}$. Both the intra-valley scattering, $W^{(++)}_{\boldsymbol{k}\boldsymbol{k}^{\prime}}$ and $W^{(--)}_{\boldsymbol{k}\boldsymbol{k}^{\prime}}$, as well as the inter-valley ones, $W^{(-+)}_{\boldsymbol{k}\boldsymbol{k}^{\prime}}$ and $W^{(+-)}_{\boldsymbol{k}\boldsymbol{k}^{\prime}}$, are included. TRS requires that $W^{(++)}_{\boldsymbol{k}\boldsymbol{k}^{\prime}}=W^{(--)}_{\boldsymbol{k}'\boldsymbol{k}}$ and $W^{(-+)}_{\boldsymbol{k}\boldsymbol{k}^{\prime}}=W^{(+-)}_{\boldsymbol{k}'\boldsymbol{k}}$. Here, we assume detailed balance, which follows from the first Born approximation. 

We look for a static solution of the SBE, Eq.~\eqref{Eq:SBE0}, within linear response. The distribution function can be conveniently parameterized by harmonic coefficients:
\begin{widetext}
  \begin{equation}
    f^{(s)}_{\boldsymbol{k}}=f_0+x_0^{(s)}\frac{\partial f_0}{\partial\epsilon}+\sum_{n=1}^{\infty} x_n^{(s)}\cos n\theta_{\kv}\frac{\partial f_0}{\partial\epsilon}+\sum_{n=1}^{\infty} y_n^{(s)}\sin n\theta_{\kv}\frac{\partial f_0}{\partial\epsilon},
  \end{equation}
\end{widetext}
where $f_0$ is the equilibrium Fermi distribution function, the angle $\theta_{\kv}$ is defined for each valley as shown in Fig.~\ref{Fig:SimplifiedScattering}.
The harmonic coefficients, $x^{(s)}_n$ and $y^{(s)}_n$, are proportional to  $\boldsymbol{E}$ within linear response. The valley density difference is determined by:
\begin{align}
  \Delta n_0 =&n^{(+)}-n^{(-)}=\int \diff \Gamma\left(f^{(+)}_{\boldsymbol{k}}-f^{(-)}_{\boldsymbol{k}}\right)\non\\
  =&-\nu\left(x^{(+)}_0-x^{(-)}_0\right),
     \label{Eq:VDDandZeroHarmonic}
\end{align}
where $\nu$ is the density of states at the Fermi level. Notice that the simplification to the second line of Eq.~(\ref{Eq:VDDandZeroHarmonic}) is a result of the assumed circular Fermi pockets, see Fig.~\ref{Fig:SimplifiedScattering}.

Before presenting the solution of the Boltzmann equation, we point out
the central role of inter-valley scattering. Namely, to obtain
non-zero valley density difference, $\Delta n_0$, the inter-valley
scattering rate must be treated with care. In particular, a constant
inter-valley scattering rate $1/\tau'$ cannot generate a nonzero
valley density difference in the static limit. To see this point, one
can integrate the SBE, Eq.~(\ref{Eq:SBE0}), over the full Brillouin
zone, assuming that the inter-valley scattering rate is a constant,
$W^{\text{inter}}$. The result is a continuity equation for the valley
densities:
$\partial_t
n^{(\pm)}-\nabla\cdot\boldsymbol{j}^{(\pm)}=-\left(n^{(\pm)}-n^{(\mp)}\right)/\tau^{\prime}$, where the
inter-valley scattering time is defined as
$\tau^{\prime-1}=\nu W^{\text{inter}}$.  In a spatially uniform and time independent system, the right hand side must vanish, which indicates that the valley density
  difference always relaxes and vanishes in the static limit, even though we allowed for non-zero (but constant) inter-valley scattering.  To avoid this problem, one must account for momentum-dependent inter-valley scattering, which will induce a ``source'' in the continuity equation.  

Following the reasoning above, we consider the inter-valley scattering rate given by:
\begin{equation}
  \begin{split}
    &W^{(-+)}_{\boldsymbol{k}\boldsymbol{k}^{\prime}}=W^{(+-)}_{\boldsymbol{k}^{\prime}\boldsymbol{k}}=\frac{1}{\nu\tau^{\prime}}\left(1+a_1\cos\theta_{\kv}+b_1\sin\theta_{\kv}\right.\\
    &\ \ \ \ \ \ \ \ \ \ \ \ \ \ \ \ \ \ \ \ \ \ \ \ \ \ \ \ \ \ \ \ \ \  \left.+a_1^{\prime}\cos\theta_{\kv'}+b_1^{\prime}\sin\theta_{\kv'}\right)\\
  \end{split}
  \label{Eq:InterValleyScattering}
\end{equation}
which explicitly breaks rotational symmetry. The dimensionless
parameters $a_1^{\vphantom\prime},a_{1}^{\prime}$ and
$b_1^{\vphantom\prime},b_{1}^{\prime}$ are determined by the
microscopic mechanisms of breaking rotational symmetry.

For demonstration purposes, we make two additional simplifications. First, we assume the intra-valley scattering is constant,
\begin{equation}
  W^{(++)}_{\boldsymbol{k}\boldsymbol{k}^{\prime}}=W^{(--)}_{\boldsymbol{k}\boldsymbol{k}^{\prime}}=\frac{1}{\nu\tau},
\end{equation}
where $\nu$ and $\tau$ are the density of states and the intra-valley
scattering time, respectively. Second, the Fermi surfaces are assumed
to be circular. Indeed, given the inter-valley scattering rate in
Eq.~(\ref{Eq:InterValleyScattering}), the detailed form of the
intra-valley scattering and the Fermi surface geometry are expected to
play a secondary role on the generation of valley polarization. They
do not affect whether a valley density difference can be generated
by external bias or not. They only affect the magnitude of the valley
density difference, at a similar level to other microscopic details that are beyond our model calculations.

The solution to Eq.~\eqref{Eq:SBE0} is physically intuitive in the
limit that the intra-valley scattering time ($\tau$) is much shorter
than the inter-valley one ($\tau'$).  To  leading order in
$\tau/\tau'$, the static solution satisfies the SBE with \emph{only}
the intra-valley scattering,
\begin{align}
  e\boldsymbol{E}\cdot\boldsymbol{v}^{(s)}_{\boldsymbol{k}}\frac{\partial f_0}{\partial\epsilon}=I_{\text{intra}}^{(s)}[f_{\boldsymbol{k}}].
\end{align}
The harmonic expansion coefficients of the distribution function are
\begin{align}
  \begin{cases}
    x_n^{(s)}=0 \ \ \ \ \ \ \ \ \ \ \ \  y_n^{(s)}=0 & \text{if $n\neq 0, 1$}\\
    x_1^{(s)}=s\, e E_x \vF \tau \ \  y_1^{(s)}= s\, e E_y \vF \tau 
  \end{cases},
  \label{Eq:IntraSol}
\end{align}
where $\tau$ is the intra-valley scattering time, $\vF$ is the Fermi velocity defined from $\vel_{\boldsymbol{k}_F}^{(s)}=s\, \vF (\cos\theta_{\kv},\sin\theta_{\kv})$. Note that without inter-valley scattering, there is no constraint on $x^{(s)}_0$ from the SBE because the number density of each valley is separately conserved. 

Now the static valley density difference can be determined by solving $0= I_{\text{inter}}^{(s)}[f_{\boldsymbol{k}}]$:
\begin{align}
  0=x^{(+)}_0-x^{(-)}_0+\frac{1}{2}\left(a_1^{\prime}x^{(+)}_1+b_1^{\prime}y^{(+)}_1-a_1x^{(-)}_1-b_1y^{(-)}_1\right).
  \label{eq:intervallySBE}
\end{align}
This equation dictates a balance between the inter-valley relaxation process (the first two terms) and a ``source'' (the last term in parentheses) that generates the valley density difference. The ``source'' originates from the interplay between the nonequilibrium distribution function from Eq.~(\ref{Eq:IntraSol}) and the rotational symmetry breaking of the inter-valley scattering rate, Eq.~(\ref{Eq:InterValleyScattering}).

Solving Eq.~(\ref{eq:intervallySBE}), we find that the valley density difference $\Delta n_0=n^{\scriptscriptstyle (+)}-n^{\scriptscriptstyle (-)}$ is given by:
          %           Substituting in the solution of $x^{(s)}_1$ and $y^{(s)}_1$, one may obtain the valley polarization as:
% \begin{equation*}
%   \resizebox{0.48\textwidth}{!}{$\Delta n_0=n^{(+)}-n^{(-)}=\frac{1}{2}\nu v_{\text{F}}\tau \left[eE_x\left(a_1+a_1^{\prime}\right)+ eE_y\left(b_1+b_1^{\prime}\right)\right],$}
% \end{equation*}
\begin{equation}
  \Delta n_0=\frac{\nu v_{\text{F}}\tau}{2} \left[eE_x\left(a_1\!+\!a_1^{\prime}\right)+ eE_y\left(b_1\!+\!b_1^{\prime}\right)\right],
\end{equation}
or, equivalently expressed in terms of the current density $\boldsymbol{j}=\sigma\boldsymbol{E}$:
\begin{equation}
  \Delta n_0=\frac{1}{2hv_{\text{F}}}\frac{h}{e^2}\left[ej_x\left(a_1+a_1^{\prime}\right)+ej_y\left(b_1+b_1^{\prime}\right)\right].
  \label{Eq:CurrentInducedVP}
\end{equation}
Notice the bulk longitudinal conductivity $\sigma$ is related to the intra-valley scattering rate $\tau$ through $\sigma=2\frac{e^2}{h}h\nu D$,
where $D=\frac{1}{2}v_{\text{F}}^2\tau$ is the two dimensional diffusion constant and the prefactor of $2$ accounts for the two valleys.

\begin{figure}[tb]
  \centering
  \includegraphics[scale=0.55]{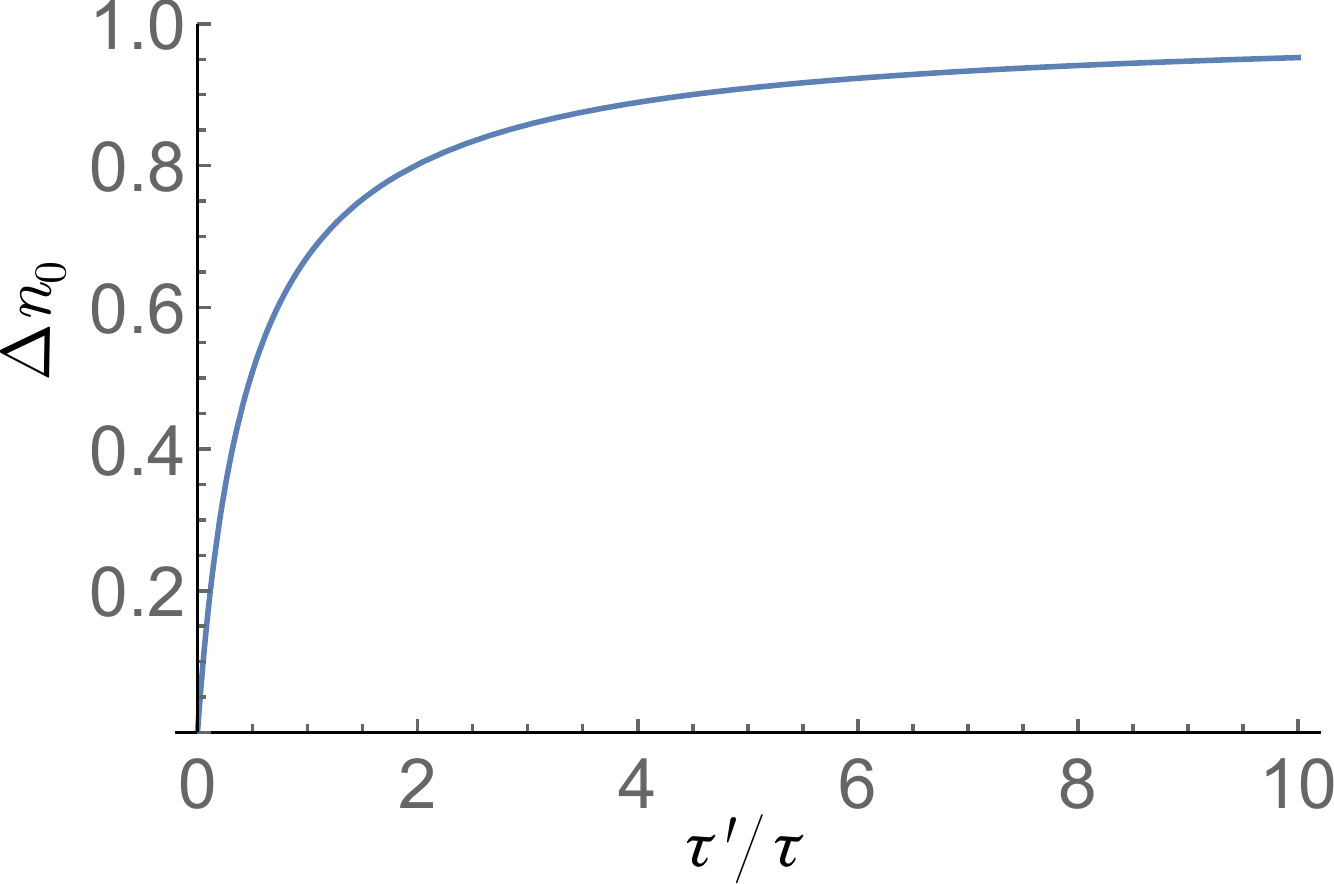}
  \caption{The valley density difference, $\Delta n_0$, as a function of the ratio of the inter- and intra valley scattering time, $\tau^{\prime}/\tau$. The density is normalized to the value of Eq.~(\ref{Eq:CurrentInducedVP}).}
  \label{Fig:ToyModelValleyDensityDifference}
\end{figure}

The simplified model presented in this subsection can be solved exactly. The valley density difference for a general ratio of inter- and intra-valley scattering time, $\tau^{\prime}/\tau$, is shown in Fig.~\ref{Fig:ToyModelValleyDensityDifference}. Indeed, when the inter-valley scattering time is much longer than the intra-valley one, the valley density difference saturates to a value given by Eq.~(\ref{Eq:CurrentInducedVP}). On the other hand, $\Delta n_0$ decreases with decreasing inter-valley scattering time. $\Delta n_0$ vanishes when the inter-valley relaxation time $\tau^{\prime}$ goes to zero.

As has been emphasized,  inter-valley scattering is essential to obtain a current induced valley density difference, because it is the channel to exchange electrons between the two valleys. Without inter-valley scattering, the electron density within each valley is exactly conserved. 

The valley density difference in Eq.~\eqref{Eq:CurrentInducedVP} is determined by the first harmonic of the inter-valley scattering rate, which explicitly breaks the discrete rotational symmetry of the system to $\mathcal{C}_{1z}$. In the next section, we determine the coefficients $a_1, a'_1, b_1, b'_1$ in Eq.~\eqref{Eq:InterValleyScattering} from microscopic modeling of h-BN aligned TBG with $\mathcal{C}_{1z}$ symmetry. 

Finally, the valley density difference is proportional to the current, Eq.~(\ref{Eq:CurrentInducedVP}), as we restricted ourselves to linear response. By reversing the current direction, the valley density difference is also reversed, and hence so is the valley polarization, see Eq.~(\ref{eq:VPOPsol}). Therefore, we conclude that with  broken rotational symmetry, the valley polarization can be  controlled by a DC current.

\subsection{Twisted Bilayer Graphene}

In this subsection, we estimate the valley density difference for the TBG/h-BN system. As has been emphasized, to induce $\Delta n_0\neq 0$ from a bias electric field, the lattice rotational symmetry needs to be fully broken. While unaligned TBG exhibits the higher symmetry point group  $D_3$, a close alignment of either top or bottom TBG layer with h-BN not only breaks the sublattice (inversion) symmetry, but also can induce strain to the sample that further breaks $\mathcal{C}_{3z}$ to $\mathcal{C}_{1z}$. In the following, we model the rotational symmetry breaking by strain. 

A full account of the microscopic details of magic angle TBG to obtain
the valley density difference is quite challenging, and requires the
full knowledge of the inter-valley scattering mechanism as well as the
spectrum and wavefunctions of TBG near the magic twist
angle. Nevertheless, the mechanism we proposed in
Sec.~\ref{subsec:SBE} is generic. The magnitude of the effect,
parametrized by the coefficient $\jcoeff$ in the valley density
difference Eq.~\eqref{Eq:ValleyDensityDifferenceGeneral}, is a
reflection of the degree of rotational symmetry breaking. For example,
in strained single layer graphene, $\jcoeff\sim \epsilon$, where
$\epsilon$ is the strain strength. In this sense, a general mechanism
that enhances the effect of strain would be desirable to explain the
small critical current observed in the experiment. In the following, we show
that in TBG aligned with h-BN, due to the interplay between two
comparable lengths --  the moir\'e scale ($a/\theta_w$) and the strain
scale ($a/\epsilon$) -- the strain effect is enhanced to
$\epsilon\rightarrow \frac{\epsilon}{\theta_w}$. To demonstrate this
point, it is enough to introduce the inter-layer tunneling
perturbatively, which preserves the analytical solubility.

\begin{figure}[tb]
  \centering
  \includegraphics[width=\linewidth]{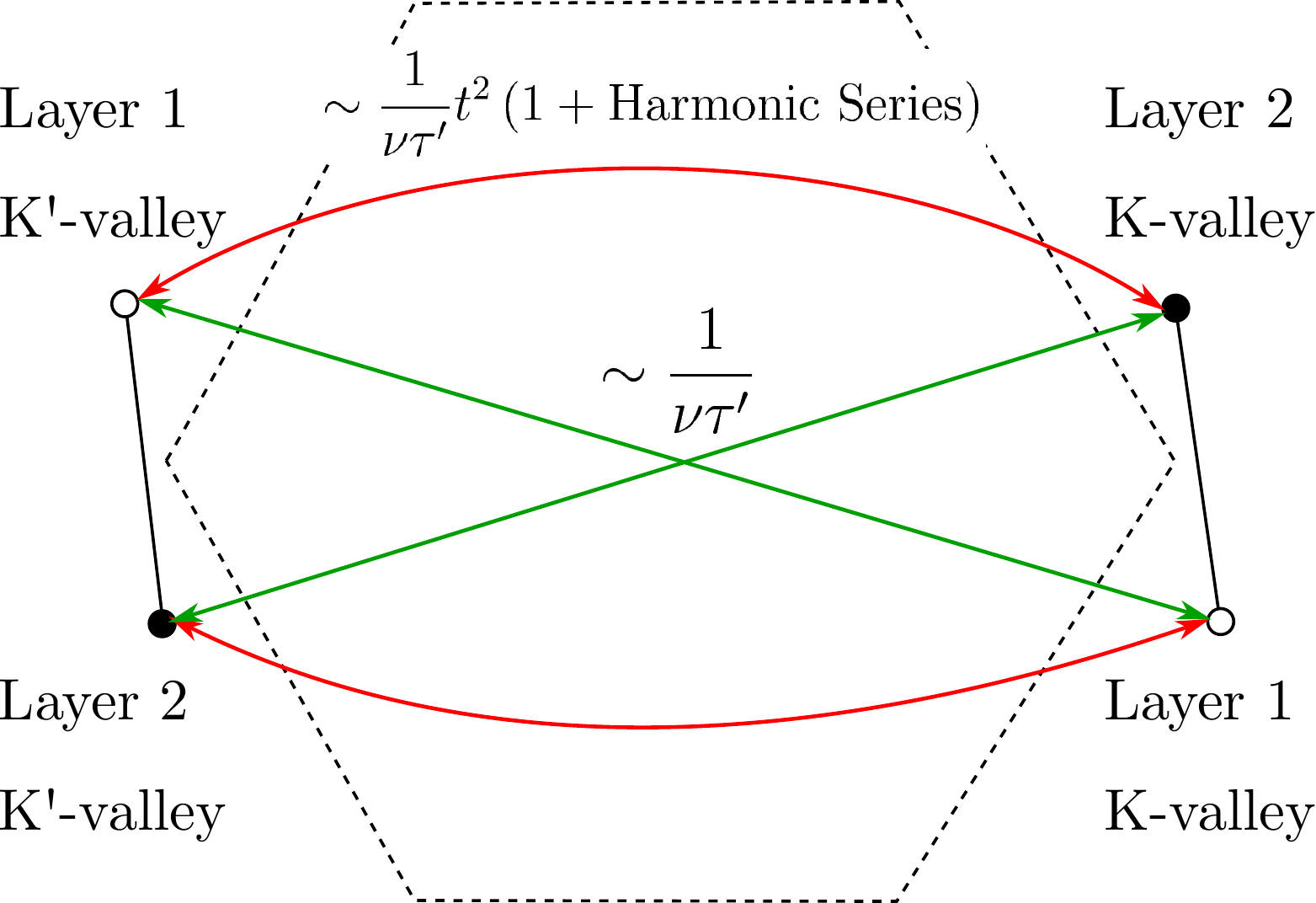}
  \caption{The inter-valley impurity scattering in twisted bilayer graphene. Layer 1 is strained, while layer 2 is not. The intra-layer inter-valley scattering (green arrows) are assumed to be isotropic. The inter-layer inter-valley scattering (red arrows) may be anisotropic, as in Eq.~(\ref{Eq:InterValleyScattering}) and Eq.~(\ref{Eq:InterVInterLScattering}).}
  \label{Fig:TBLGScattering}
\end{figure}

Our modeling is based on the continuous model, introduced in Ref.~\onlinecite{bistritzer2011moire} and generalized in Ref.~\onlinecite{balents2019general} that captures elastic deformations systematically.	Here, we assume a uniaxial strain parameterized by the strain tensor\cite{he2020giant}
\begin{align}
  \straintensor=-\frac{(1+\nue)\epsilon}{2}
  \begin{bmatrix}
    \cos2\phi & \sin 2\phi\\
    \sin 2\phi  & -\cos 2\phi
  \end{bmatrix}+\frac{(\nue-1)\epsilon}{2}\mathbb{I}_2,
                  \label{eq:strain}
\end{align} 
where $\epsilon$ is a dimensionless parameter characterizing the strain strength. $\nu_{\epsilon}=0.165$ is the Poisson ratio for graphene. $\phi$ is the direction of the strain. Note that only the first term in Eq.~\eqref{eq:strain} breaks rotational symmetry and enters into $\jcoeff$. 

Without loss of generality, we consider the strain only on layer 1, Fig.~\ref{Fig:TBLGScattering}.  Due to the combination of the strain field and the alignment with the h-BN substrate, the Dirac Hamiltonian around the K-point becomes $H=v_{\text{D}}\boldsymbol{k}\cdot(1+\boldsymbol{\mathcal{E}})\cdot\boldsymbol{\sigma}+m\sigma_z$\cite{balents2019general}, where $v_{\text{D}}$, $\boldsymbol{k}$ and $m$ are the Dirac velocity, the momentum measured from the Dirac point and the mass gap, respectively. At the leading order in the strain strength, the rotational symmetry breaking of the continuous model under the strain field is reflected in several aspects. \emph{First}, the Dirac points at $K^{(\prime)}$ valleys are shifted by $\delta \ve{K}^{(\prime)}_i=-\straintensor \cdot \ve{K}^{(\prime)}_i$ for the strained layer\cite{balents2019general}. As a result, the momentum difference between the adjacent Dirac points of the two layers, see Fig.~\ref{Fig:StrainAndBZ}, is modified as 
\begin{align}
  \ve{q}_i\rightarrow \ve{q}'_i=\ve{q}_i-\straintensor\cdot \ve{K}_i
\end{align}
where $\ve{q}_i=\theta_w \ve{K}_i\times \hat{z}$, $i\in\{1,2,3\}$.
\emph{Second}, the single layer hopping integral is modified due to
the strain field, which modifies the single layer energetics and shift
the Dirac points also at order $\epsilon/a$. \emph{Third}, the Dirac
spectrum is anisotropic. However, the Dirac spectrum anisotropy is
parametrically smaller in $|\ve{k}|/|\ve{K}|$ than the shift of the
Dirac point,~\cite{balents2019general} and is thus neglected.

For simplicity, only the first contribution is included in the
following discussions. As the wave vectors for inter-layer
tunneling $\ve{q}'_i-\ve{q}'_1$, i.e.\ the the reciprocal
lattice vector of the moir\'e Brillouin zone (mBZ), are modified due to the strain, 
                  %                   \MY{I suggest to add discussions on the modifications to the scattering structure calculation explicitly} 
the $\mathcal{C}_3$ rotational symmetry of the mBZ is broken explicitly [see Fig.~\ref{Fig:StrainAndBZ}(b)], and the effect is characterized by:
\begin{equation}
  \frac{\delta\boldsymbol{K}_\epsilon}{|\ve{q}_i|}\sim\frac{\epsilon\left|\boldsymbol{K}\right|}{\theta_w\left|\boldsymbol{K}\right|}=\frac{\epsilon}{\theta_w},
\end{equation}
where both the shift of the Dirac point under strain field, $\delta\boldsymbol{K}\sim\epsilon\left|\boldsymbol{K}\right|$, and the size of mBZ, $q=\theta_w\left|\boldsymbol{K}\right|$, are small and comparable to each other.

We  focus on bulk transport. From the discussion of Sec.~\ref{subsec:SBE}, both intra- and inter-valley scattering of TBG must be taken into account properly. The dominant scattering mechanism in TBG is yet to be determined, but certain key features may be captured by simple modeling. Here, we consider short range impurities and a low doping level (well below $3/4$ filling of the moir\'e conduction band) so that there is a Fermi pocket around each Dirac point at the corner of the mBZ (see Fig.~\ref{Fig:TBLGScattering}). Notice that the inter-valley scattering involves a much larger momentum transfer than the intra-valley one. Therefore, we assume the intra-valley scattering time $\tau$ being much shorter than the inter-valley scattering time $\tau^{\prime}$, which is expected to be generically true for most scattering mechanisms in TBG.
                  %                   Near the transition temperature, one should focus on the bulk transport. The dominant scattering mechanism in TBG is yet to be determined. For simplicity, we assume short ranged impurities. 
                  % 	%              to capture the necessary features, including the inter-valley scattering and the broken rotational symmetry.

                  %              We took into account both the intra- and inter-valley scattering for twisted bulayer graphene. For simplicity, we consider low doping level so that there are four Fermi pockets around each Dirac point (two Fermi pockets for each valley), Fig.~\ref{Fig:TBLGScattering}. Notice that the inter-valley scattering involves a much larger momentum transfer than the intra-valley one. Therefore, we assume the intra-valley scattering time $\tau$ being much shorter than the inter-valley scattering time $\tau^{\prime}$.

\begin{figure}[tb]
  \centering
  \includegraphics[width=\linewidth]{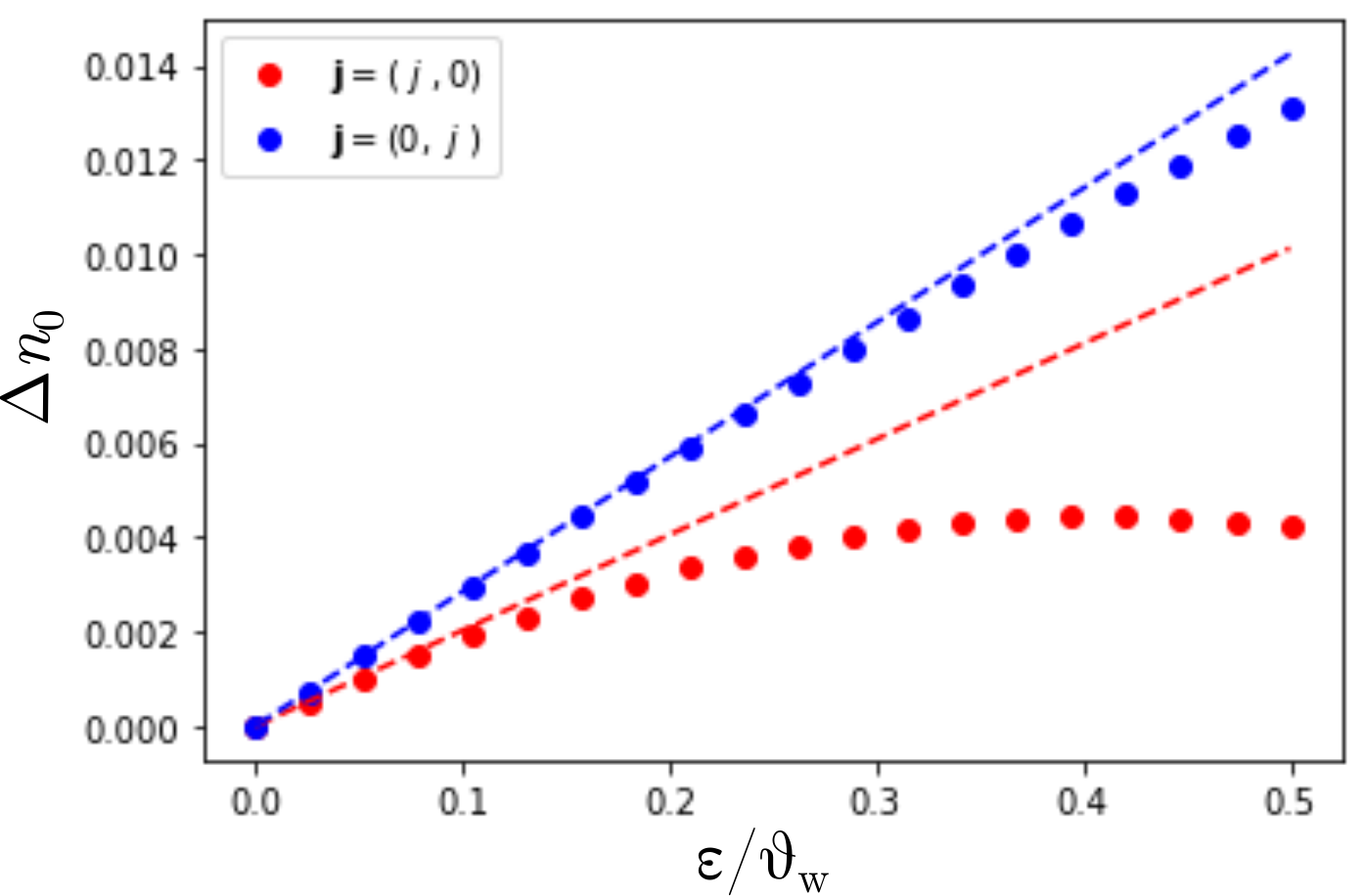}
  \caption{The valley density difference generated by a DC current in TBG with a strain field. The density is normalized to $\frac{3t^2/4}{1+3t^2/4}\frac{1}{hv_F}\frac{h}{e^2}ej$. The dotted lines are obtained from a numerical calculation. The dashed lines correspond to the analytical expression of Eq.~(\ref{Eq:TBGResult}) for comparison. The parameters are chosen as $\theta_w=1.03^{\circ}$, $\phi=17.20^{\circ}$, $v_{\text{D}}q/m=0.1$ and $v_{\text{D}}k_{\text{F}}/m=0.005$.}
  \label{Fig:result}
\end{figure}

As discussed in the previous subsection, the intra-valley scattering
only plays a secondary role in the generation of valley density
difference by a bias DC current. Hence, we assume the intra-valley
scattering is dominated by the scattering within each Fermi pocket
with a constant relaxation time, $\tau$.  The inter-valley scattering
requires more careful examination to obtain the coefficients
$a_1, a'_1, b_1, b'_1$ in Eq.~\eqref{Eq:InterValleyScattering}. There
are several processes as shown in Fig.~\ref{Fig:TBLGScattering}. With
simple on-site disorder, it turns out that the scattering between the
two valleys in the same graphene layer does not break $\mac{C}_{3z}$
and generate non-zero $a_1, a'_1, b_1, b'_1$
% have the necessary complicated structure, 
since we neglected the anisotropy of the Dirac spectrum. Thus, the scattering rates between the two valleys of the same graphene layer are taken to be constant, $\left(\nu\tau^{\prime}\right)^{-1}$, as indicated by the green arrows in Fig.~\ref{Fig:TBLGScattering}. At the same time, the scattering rates between the two valleys in different layers may break rotational symmetry through the process indicated by the red arrows in Fig.~\ref{Fig:TBLGScattering}. 
% The complicated scattering rates 
It is directly related to the shift of the Dirac points due to strain as well as the coherent inter-layer tunneling. As in Fig.~\ref{Fig:TBLGScattering}, the scattering rates of such processes are of order $\sim\left(\nu\tau^{\prime}\right)^{-1}t^2$, where $t$ is a dimensionless perturbation parameter for the inter-layer tunneling, Eq.~(\ref{Eq:WeakTunneling}).

After fitting the scattering rates into the semi-classical Boltzmann equation introduced in the previous subsection, we obtain the valley density difference generated by a DC current. The main result is summarized in Fig.~\ref{Fig:result}, where, without loss of generality, only one graphene layer is strained (see Appendix~\ref{Sec:ModelingofTBG} for more details of the calculation).
                  %                   This result is obtained by straining only one of the graphene layers, without the loss of generality.

When the strain strength is zero, $\mac{C}_{3z}$ is restored and the valley density difference is identically zero.
                  %                   due to the $\mathcal{C}_3$ rotational symmetry of the moir\'e lattice in the absence of strain field. 
For non-zero strain strength, the valley density difference appears as expected from the breaking of $\mathcal{C}_3$ symmetry. At small strain strength, the valley density difference is linear in the strain strength:
\begin{equation}
  \frac{\Delta n_0}{n_{\ve{j}}} = 6(1+\nu_{\epsilon})\frac{v_{\text{D}}k_{\text{F}}}{m}\frac{\epsilon}{\theta_w}\sin{(2\phi+\theta_{\ve{j}})},
  \label{Eq:TBGResult}
\end{equation}
                  %                   \begin{equation}
                  %                     \Delta n_0 = 6(1+\nu_{\epsilon})\frac{k}{m}\frac{\epsilon}{\theta_w}\times\left\{\begin{split}
                  %           &\sin 2\phi\ \ \ \ \boldsymbol{j}=(j,0)\\
          %               &\cos 2\phi\ \ \ \ \boldsymbol{j}=(0,j)
                            %                             \end{split}\right.
                            %                             \label{Eq:TBGResult}
                            %                             \end{equation}
where the basic scale for current induced density is
$n_{\ve{j}}=\frac{3t^2/4}{1+3t^2/4}\frac{1}{hv_F}\frac{h}{e^2}ej$. Here
$\theta_{\ve{j}}$ is the angle of the current, $\phi$ is the angle of
uniaxial strain as defined in Eq.~\eqref{eq:strain}. The small
numerical value in Fig.~\ref{Fig:result} is a result of the low doping
assumption, where $v_{\text{D}}k_{\text{F}}/m$ is a small parameter with $k_{\text{F}}$ being the Fermi momentum. The result is further suppressed by the inter-layer
tunneling $t^2$, which is assumed to be small to introduce the
inter-layer tunneling perturbatively.

Even though our result, Fig.~\ref{Fig:result} and Eq.~(\ref{Eq:TBGResult}), was obtained with a set of very specific assumptions (weak inter-layer coupling and low doping level), some implications can be drawn for  real samples. In reality, there are several comparable scale: moir\'e band width, inter-layer coupling and the mass gap\cite{bistritzer2011moire,kim2018accurate}. Therefore, the dimensionless parameter $t^2$ is not small. In addition, the phenomenon of current switching of valley polarization is observed at three quarter filling. Thus, it is reasonable to expect that the small factor of $v_{\text{D}}k_{\text{F}}/m$ in Eq.~(\ref{Eq:TBGResult}) is lifted and is on the order of $\mathcal{O}(1)$. Therefore, we conclude that the actual valley density difference can be estimated as: 
\begin{equation}
  \Delta n_0\propto \frac{\epsilon}{\theta_w}\frac{1}{hv_F}\frac{h}{e^2}ej
\end{equation}
with a numerical coefficient on the order of $\mathcal{O}(1)$. Based on the arguments above, one may roughly estimate that a small DC current ($\sim 10 ^{-3} \text{A}/\text{m}$) could generate a large valley density difference on the order of $10^{11}\text{m}^{-2}$. This is comparable to the effect of magnetic field ($\sim \nu\mu_{\text{B}}B$ and $B\sim 10\ \text{mT}$ with $\mu_{\text{B}}$ being the Bohr magneton)\cite{serlin2020intrinsic}.

\section{Summary and discussion}
In this work, we proposed a mechanism for DC current switching of the
valley polarization in the dissipative regime.  It was obtained by
first determining the dynamics of the valley polarization order
parameter (VPOP) in an applied electric field near the Curie
temperature $T_c$, using the nonequilibrium Keldysh
formalism.  This formalism relates the spontaneous value of the VPOP to the one linearly induced by a current without interactions in the paramagnetic phase.  In this way, one sees that  sweeping the DC current, and thus varying the current
generated valley density difference, the VPOP undergoes a first order
phase transition.   Consequently, the model reproduces  a hysteresis curve
in the Hall conductance, consistent  with the experiments in
Refs.~\cite{serlin2020intrinsic,sharpe2019emergent}. The current
generated valley density difference takes the generic form of
Eq.~\eqref{Eq:ValleyDensityDifferenceGeneral} and we repeat here
\begin{align}
  \Delta n_0 \simeq \frac{1}{e v_\text{F}} \ve{j}\cdot \boldsymbol{\delta}_\epsilon.
\end{align}
From a solution of the semi-classical Boltzmann equation, we point out
that a proper form of inter-valley scattering that breaks the
$\mac{C}_{3z}$ rotational symmetry is necessary to generate non-zero
valley density difference by the transport current.  This may be due
to strain in TBG aligned with h-BN. Our modeling indicates an
enhancement of the strain effect by a factor inversely proportional of the twist angle,
i.e. $|\boldsymbol{\delta}_\epsilon|\sim \epsilon/\theta_w$. Together
with the reduction of Fermi velocity in magic angle TBG, we argue
that these two effects significantly reduce the critical current needed to
reverse the Hall conductance.

A few theories have been developed in light of the observation of
current switching of valley polarization in the TBG/h-BN
sample~\cite{serlin2020intrinsic,he2020giant}.  The picture developed
here relies on the finite scattering time $\tau$, and thus does not
apply to the quantum anomalous Hall regime observed at temperature
well below $T_c$.  It is also a bulk mechanism.  The theoretical analysis in Ref.~[\onlinecite{serlin2020intrinsic}], in contrast, describes a finite-size mechanism which applies in the non-dissipative
limit $\sigma_{xx}\ll\sigma_{xy}$, based on edge states.  In that
limit, each edge state can be describes in quasi-equilibrium even in
the presence of a current, and thereby
Ref.~\onlinecite{serlin2020intrinsic}\ obtain s a correction to the
edge state Free energy of order $I^3$, where $I$ is the edge current,
which favors the valley polarization associated with a particular
direction of the edge current.  Another treatment in the
non-dissipative regime explicitly models the forces on a domain wall,
Ref.~\onlinecite{MacDoanld2020}.  In connection to the present work,
we note that Ref.~\onlinecite{MacDoanld2020}\ introduces violation of
valley conservation in a phenomenological manner.  At intermediate
temperature below $T_c$, the system has non-zero valley polarization
and orbital magnetization and is in the dissipative transport regime.
Ref.~[\onlinecite{he2020giant}] presents a mechanism for current
reversal of the anomalous Hall effect due to magnetoelectric response.
The latter work does not distinguish valley polarization and orbital
magnetization, which play very different roles in our treatment due to
the quasi-conservation of the former.  In any case, the result quoted
in Ref.~[\onlinecite{he2020giant}] becomes negligible close to
$T_c$. Our work instead is relevant near $T_c$, when the orbital
magnetization of the equilibrium system is too small to be greatly
affected by a small critical DC current.

\section{Acknowledgement}
We thank Kohei Kawabata for helpful discussions. This research is funded in part by the Gordon and Betty Moore Foundation through Grant GBMF8690 to UCSB to support the work of M.Y.  L.B. is supported by the NSF CMMT program under Grant No. DMR-1818533. X.Y. is supported by the Heising-Simons Foundation, the Simons Foundation, and NSF Grant No. NSF PHY-1748958 and partly by NSF Grant No. DMR-1608238 and DMR-2037654.

\bibliography{CSVPTBG_Ref_updated}{}	

\begin{thebibliography}{43}%
\makeatletter
\providecommand \@ifxundefined [1]{%
 \@ifx{#1\undefined}
}%
\providecommand \@ifnum [1]{%
 \ifnum #1\expandafter \@firstoftwo
 \else \expandafter \@secondoftwo
 \fi
}%
\providecommand \@ifx [1]{%
 \ifx #1\expandafter \@firstoftwo
 \else \expandafter \@secondoftwo
 \fi
}%
\providecommand \natexlab [1]{#1}%
\providecommand \enquote  [1]{``#1''}%
\providecommand \bibnamefont  [1]{#1}%
\providecommand \bibfnamefont [1]{#1}%
\providecommand \citenamefont [1]{#1}%
\providecommand \href@noop [0]{\@secondoftwo}%
\providecommand \href [0]{\begingroup \@sanitize@url \@href}%
\providecommand \@href[1]{\@@startlink{#1}\@@href}%
\providecommand \@@href[1]{\endgroup#1\@@endlink}%
\providecommand \@sanitize@url [0]{\catcode `\\12\catcode `\$12\catcode
  `\&12\catcode `\#12\catcode `\^12\catcode `\_12\catcode `\%12\relax}%
\providecommand \@@startlink[1]{}%
\providecommand \@@endlink[0]{}%
\providecommand \url  [0]{\begingroup\@sanitize@url \@url }%
\providecommand \@url [1]{\endgroup\@href {#1}{\urlprefix }}%
\providecommand \urlprefix  [0]{URL }%
\providecommand \Eprint [0]{\href }%
\providecommand \doibase [0]{http://dx.doi.org/}%
\providecommand \selectlanguage [0]{\@gobble}%
\providecommand \bibinfo  [0]{\@secondoftwo}%
\providecommand \bibfield  [0]{\@secondoftwo}%
\providecommand \translation [1]{[#1]}%
\providecommand \BibitemOpen [0]{}%
\providecommand \bibitemStop [0]{}%
\providecommand \bibitemNoStop [0]{.\EOS\space}%
\providecommand \EOS [0]{\spacefactor3000\relax}%
\providecommand \BibitemShut  [1]{\csname bibitem#1\endcsname}%
\let\auto@bib@innerbib\@empty
%</preamble>
\bibitem [{\citenamefont {Serlin}\ \emph {et~al.}(2020)\citenamefont {Serlin},
  \citenamefont {Tschirhart}, \citenamefont {Polshyn}, \citenamefont {Zhang},
  \citenamefont {Zhu}, \citenamefont {Watanabe}, \citenamefont {Taniguchi},
  \citenamefont {Balents},\ and\ \citenamefont {Young}}]{serlin2020intrinsic}%
  \BibitemOpen
  \bibfield  {author} {\bibinfo {author} {\bibfnamefont {M.}~\bibnamefont
  {Serlin}}, \bibinfo {author} {\bibfnamefont {C.}~\bibnamefont {Tschirhart}},
  \bibinfo {author} {\bibfnamefont {H.}~\bibnamefont {Polshyn}}, \bibinfo
  {author} {\bibfnamefont {Y.}~\bibnamefont {Zhang}}, \bibinfo {author}
  {\bibfnamefont {J.}~\bibnamefont {Zhu}}, \bibinfo {author} {\bibfnamefont
  {K.}~\bibnamefont {Watanabe}}, \bibinfo {author} {\bibfnamefont
  {T.}~\bibnamefont {Taniguchi}}, \bibinfo {author} {\bibfnamefont
  {L.}~\bibnamefont {Balents}}, \ and\ \bibinfo {author} {\bibfnamefont
  {A.}~\bibnamefont {Young}},\ }\href@noop {} {\bibfield  {journal} {\bibinfo
  {journal} {Science}\ }\textbf {\bibinfo {volume} {367}},\ \bibinfo {pages}
  {900} (\bibinfo {year} {2020})}\BibitemShut {NoStop}%
\bibitem [{\citenamefont {Sharpe}\ \emph {et~al.}(2019)\citenamefont {Sharpe},
  \citenamefont {Fox}, \citenamefont {Barnard}, \citenamefont {Finney},
  \citenamefont {Watanabe}, \citenamefont {Taniguchi}, \citenamefont
  {Kastner},\ and\ \citenamefont {Goldhaber-Gordon}}]{sharpe2019emergent}%
  \BibitemOpen
  \bibfield  {author} {\bibinfo {author} {\bibfnamefont {A.~L.}\ \bibnamefont
  {Sharpe}}, \bibinfo {author} {\bibfnamefont {E.~J.}\ \bibnamefont {Fox}},
  \bibinfo {author} {\bibfnamefont {A.~W.}\ \bibnamefont {Barnard}}, \bibinfo
  {author} {\bibfnamefont {J.}~\bibnamefont {Finney}}, \bibinfo {author}
  {\bibfnamefont {K.}~\bibnamefont {Watanabe}}, \bibinfo {author}
  {\bibfnamefont {T.}~\bibnamefont {Taniguchi}}, \bibinfo {author}
  {\bibfnamefont {M.}~\bibnamefont {Kastner}}, \ and\ \bibinfo {author}
  {\bibfnamefont {D.}~\bibnamefont {Goldhaber-Gordon}},\ }\href@noop {}
  {\bibfield  {journal} {\bibinfo  {journal} {Science}\ }\textbf {\bibinfo
  {volume} {365}},\ \bibinfo {pages} {605} (\bibinfo {year}
  {2019})}\BibitemShut {NoStop}%
\bibitem [{\citenamefont {Stoner}(1938)}]{stoner1938collective}%
  \BibitemOpen
  \bibfield  {author} {\bibinfo {author} {\bibfnamefont {E.~C.}\ \bibnamefont
  {Stoner}},\ }\href@noop {} {\bibfield  {journal} {\bibinfo  {journal}
  {Proceedings of the Royal Society of London. Series A. Mathematical and
  Physical Sciences}\ }\textbf {\bibinfo {volume} {165}},\ \bibinfo {pages}
  {372} (\bibinfo {year} {1938})}\BibitemShut {NoStop}%
\bibitem [{\citenamefont {Sondhi}\ \emph {et~al.}(1993)\citenamefont {Sondhi},
  \citenamefont {Karlhede}, \citenamefont {Kivelson},\ and\ \citenamefont
  {Rezayi}}]{sondhi1993skyrmions}%
  \BibitemOpen
  \bibfield  {author} {\bibinfo {author} {\bibfnamefont {S.~L.}\ \bibnamefont
  {Sondhi}}, \bibinfo {author} {\bibfnamefont {A.}~\bibnamefont {Karlhede}},
  \bibinfo {author} {\bibfnamefont {S.~A.}\ \bibnamefont {Kivelson}}, \ and\
  \bibinfo {author} {\bibfnamefont {E.~H.}\ \bibnamefont {Rezayi}},\ }\href
  {\doibase 10.1103/PhysRevB.47.16419} {\bibfield  {journal} {\bibinfo
  {journal} {Phys. Rev. B}\ }\textbf {\bibinfo {volume} {47}},\ \bibinfo
  {pages} {16419} (\bibinfo {year} {1993})}\BibitemShut {NoStop}%
\bibitem [{\citenamefont {MacDonald}\ \emph {et~al.}(1996)\citenamefont
  {MacDonald}, \citenamefont {Fertig},\ and\ \citenamefont
  {Brey}}]{macdonald1996skyrmions}%
  \BibitemOpen
  \bibfield  {author} {\bibinfo {author} {\bibfnamefont {A.~H.}\ \bibnamefont
  {MacDonald}}, \bibinfo {author} {\bibfnamefont {H.~A.}\ \bibnamefont
  {Fertig}}, \ and\ \bibinfo {author} {\bibfnamefont {L.}~\bibnamefont
  {Brey}},\ }\href {\doibase 10.1103/PhysRevLett.76.2153} {\bibfield  {journal}
  {\bibinfo  {journal} {Phys. Rev. Lett.}\ }\textbf {\bibinfo {volume} {76}},\
  \bibinfo {pages} {2153} (\bibinfo {year} {1996})}\BibitemShut {NoStop}%
\bibitem [{\citenamefont {Cao}\ \emph {et~al.}(2018{\natexlab{a}})\citenamefont
  {Cao}, \citenamefont {Fatemi}, \citenamefont {Demir}, \citenamefont {Fang},
  \citenamefont {Tomarken}, \citenamefont {Luo}, \citenamefont
  {Sanchez-Yamagishi}, \citenamefont {Watanabe}, \citenamefont {Taniguchi},
  \citenamefont {Kaxiras} \emph {et~al.}}]{cao2018correlated}%
  \BibitemOpen
  \bibfield  {author} {\bibinfo {author} {\bibfnamefont {Y.}~\bibnamefont
  {Cao}}, \bibinfo {author} {\bibfnamefont {V.}~\bibnamefont {Fatemi}},
  \bibinfo {author} {\bibfnamefont {A.}~\bibnamefont {Demir}}, \bibinfo
  {author} {\bibfnamefont {S.}~\bibnamefont {Fang}}, \bibinfo {author}
  {\bibfnamefont {S.~L.}\ \bibnamefont {Tomarken}}, \bibinfo {author}
  {\bibfnamefont {J.~Y.}\ \bibnamefont {Luo}}, \bibinfo {author} {\bibfnamefont
  {J.~D.}\ \bibnamefont {Sanchez-Yamagishi}}, \bibinfo {author} {\bibfnamefont
  {K.}~\bibnamefont {Watanabe}}, \bibinfo {author} {\bibfnamefont
  {T.}~\bibnamefont {Taniguchi}}, \bibinfo {author} {\bibfnamefont
  {E.}~\bibnamefont {Kaxiras}},  \emph {et~al.},\ }\href@noop {} {\bibfield
  {journal} {\bibinfo  {journal} {Nature}\ }\textbf {\bibinfo {volume} {556}},\
  \bibinfo {pages} {80} (\bibinfo {year} {2018}{\natexlab{a}})}\BibitemShut
  {NoStop}%
\bibitem [{\citenamefont {Cao}\ \emph {et~al.}(2018{\natexlab{b}})\citenamefont
  {Cao}, \citenamefont {Fatemi}, \citenamefont {Fang}, \citenamefont
  {Watanabe}, \citenamefont {Taniguchi}, \citenamefont {Kaxiras},\ and\
  \citenamefont {Jarillo-Herrero}}]{cao2018unconventional}%
  \BibitemOpen
  \bibfield  {author} {\bibinfo {author} {\bibfnamefont {Y.}~\bibnamefont
  {Cao}}, \bibinfo {author} {\bibfnamefont {V.}~\bibnamefont {Fatemi}},
  \bibinfo {author} {\bibfnamefont {S.}~\bibnamefont {Fang}}, \bibinfo {author}
  {\bibfnamefont {K.}~\bibnamefont {Watanabe}}, \bibinfo {author}
  {\bibfnamefont {T.}~\bibnamefont {Taniguchi}}, \bibinfo {author}
  {\bibfnamefont {E.}~\bibnamefont {Kaxiras}}, \ and\ \bibinfo {author}
  {\bibfnamefont {P.}~\bibnamefont {Jarillo-Herrero}},\ }\href@noop {}
  {\bibfield  {journal} {\bibinfo  {journal} {Nature}\ }\textbf {\bibinfo
  {volume} {556}},\ \bibinfo {pages} {43} (\bibinfo {year}
  {2018}{\natexlab{b}})}\BibitemShut {NoStop}%
\bibitem [{\citenamefont {Dodaro}\ \emph {et~al.}(2018)\citenamefont {Dodaro},
  \citenamefont {Kivelson}, \citenamefont {Schattner}, \citenamefont {Sun},\
  and\ \citenamefont {Wang}}]{PhysRevB.98.075154}%
  \BibitemOpen
  \bibfield  {author} {\bibinfo {author} {\bibfnamefont {J.~F.}\ \bibnamefont
  {Dodaro}}, \bibinfo {author} {\bibfnamefont {S.~A.}\ \bibnamefont
  {Kivelson}}, \bibinfo {author} {\bibfnamefont {Y.}~\bibnamefont {Schattner}},
  \bibinfo {author} {\bibfnamefont {X.~Q.}\ \bibnamefont {Sun}}, \ and\
  \bibinfo {author} {\bibfnamefont {C.}~\bibnamefont {Wang}},\ }\href {\doibase
  10.1103/PhysRevB.98.075154} {\bibfield  {journal} {\bibinfo  {journal} {Phys.
  Rev. B}\ }\textbf {\bibinfo {volume} {98}},\ \bibinfo {pages} {075154}
  (\bibinfo {year} {2018})}\BibitemShut {NoStop}%
\bibitem [{\citenamefont {Zou}\ \emph {et~al.}(2018)\citenamefont {Zou},
  \citenamefont {Po}, \citenamefont {Vishwanath},\ and\ \citenamefont
  {Senthil}}]{PhysRevB.98.085435}%
  \BibitemOpen
  \bibfield  {author} {\bibinfo {author} {\bibfnamefont {L.}~\bibnamefont
  {Zou}}, \bibinfo {author} {\bibfnamefont {H.~C.}\ \bibnamefont {Po}},
  \bibinfo {author} {\bibfnamefont {A.}~\bibnamefont {Vishwanath}}, \ and\
  \bibinfo {author} {\bibfnamefont {T.}~\bibnamefont {Senthil}},\ }\href
  {\doibase 10.1103/PhysRevB.98.085435} {\bibfield  {journal} {\bibinfo
  {journal} {Phys. Rev. B}\ }\textbf {\bibinfo {volume} {98}},\ \bibinfo
  {pages} {085435} (\bibinfo {year} {2018})}\BibitemShut {NoStop}%
\bibitem [{\citenamefont {Zhang}\ \emph
  {et~al.}(2019{\natexlab{a}})\citenamefont {Zhang}, \citenamefont {Mao},
  \citenamefont {Cao}, \citenamefont {Jarillo-Herrero},\ and\ \citenamefont
  {Senthil}}]{PhysRevB.99.075127}%
  \BibitemOpen
  \bibfield  {author} {\bibinfo {author} {\bibfnamefont {Y.-H.}\ \bibnamefont
  {Zhang}}, \bibinfo {author} {\bibfnamefont {D.}~\bibnamefont {Mao}}, \bibinfo
  {author} {\bibfnamefont {Y.}~\bibnamefont {Cao}}, \bibinfo {author}
  {\bibfnamefont {P.}~\bibnamefont {Jarillo-Herrero}}, \ and\ \bibinfo {author}
  {\bibfnamefont {T.}~\bibnamefont {Senthil}},\ }\href {\doibase
  10.1103/PhysRevB.99.075127} {\bibfield  {journal} {\bibinfo  {journal} {Phys.
  Rev. B}\ }\textbf {\bibinfo {volume} {99}},\ \bibinfo {pages} {075127}
  (\bibinfo {year} {2019}{\natexlab{a}})}\BibitemShut {NoStop}%
\bibitem [{\citenamefont {Balents}\ \emph {et~al.}(2020)\citenamefont
  {Balents}, \citenamefont {Dean}, \citenamefont {Efetov},\ and\ \citenamefont
  {Young}}]{balents2020superconductivity}%
  \BibitemOpen
  \bibfield  {author} {\bibinfo {author} {\bibfnamefont {L.}~\bibnamefont
  {Balents}}, \bibinfo {author} {\bibfnamefont {C.~R.}\ \bibnamefont {Dean}},
  \bibinfo {author} {\bibfnamefont {D.~K.}\ \bibnamefont {Efetov}}, \ and\
  \bibinfo {author} {\bibfnamefont {A.~F.}\ \bibnamefont {Young}},\ }\href@noop
  {} {\bibfield  {journal} {\bibinfo  {journal} {Nature Physics}\ ,\ \bibinfo
  {pages} {1}} (\bibinfo {year} {2020})}\BibitemShut {NoStop}%
\bibitem [{\citenamefont {Po}\ \emph {et~al.}(2018{\natexlab{a}})\citenamefont
  {Po}, \citenamefont {Watanabe},\ and\ \citenamefont
  {Vishwanath}}]{PhysRevLett.121.126402}%
  \BibitemOpen
  \bibfield  {author} {\bibinfo {author} {\bibfnamefont {H.~C.}\ \bibnamefont
  {Po}}, \bibinfo {author} {\bibfnamefont {H.}~\bibnamefont {Watanabe}}, \ and\
  \bibinfo {author} {\bibfnamefont {A.}~\bibnamefont {Vishwanath}},\ }\href
  {\doibase 10.1103/PhysRevLett.121.126402} {\bibfield  {journal} {\bibinfo
  {journal} {Phys. Rev. Lett.}\ }\textbf {\bibinfo {volume} {121}},\ \bibinfo
  {pages} {126402} (\bibinfo {year} {2018}{\natexlab{a}})}\BibitemShut
  {NoStop}%
\bibitem [{\citenamefont {Zhang}\ \emph
  {et~al.}(2019{\natexlab{b}})\citenamefont {Zhang}, \citenamefont {Mao},\ and\
  \citenamefont {Senthil}}]{Zhang2019hBN}%
  \BibitemOpen
  \bibfield  {author} {\bibinfo {author} {\bibfnamefont {Y.-H.}\ \bibnamefont
  {Zhang}}, \bibinfo {author} {\bibfnamefont {D.}~\bibnamefont {Mao}}, \ and\
  \bibinfo {author} {\bibfnamefont {T.}~\bibnamefont {Senthil}},\ }\href
  {\doibase 10.1103/PhysRevResearch.1.033126} {\bibfield  {journal} {\bibinfo
  {journal} {Phys. Rev. Research}\ }\textbf {\bibinfo {volume} {1}},\ \bibinfo
  {pages} {033126} (\bibinfo {year} {2019}{\natexlab{b}})}\BibitemShut
  {NoStop}%
\bibitem [{\citenamefont {Isobe}\ \emph {et~al.}(2018)\citenamefont {Isobe},
  \citenamefont {Yuan},\ and\ \citenamefont {Fu}}]{PhysRevX.8.041041}%
  \BibitemOpen
  \bibfield  {author} {\bibinfo {author} {\bibfnamefont {H.}~\bibnamefont
  {Isobe}}, \bibinfo {author} {\bibfnamefont {N.~F.~Q.}\ \bibnamefont {Yuan}},
  \ and\ \bibinfo {author} {\bibfnamefont {L.}~\bibnamefont {Fu}},\ }\href
  {\doibase 10.1103/PhysRevX.8.041041} {\bibfield  {journal} {\bibinfo
  {journal} {Phys. Rev. X}\ }\textbf {\bibinfo {volume} {8}},\ \bibinfo {pages}
  {041041} (\bibinfo {year} {2018})}\BibitemShut {NoStop}%
\bibitem [{\citenamefont {Nandkishore}\ \emph {et~al.}(2012)\citenamefont
  {Nandkishore}, \citenamefont {Levitov},\ and\ \citenamefont
  {Chubukov}}]{nandkishore2012chiral}%
  \BibitemOpen
  \bibfield  {author} {\bibinfo {author} {\bibfnamefont {R.}~\bibnamefont
  {Nandkishore}}, \bibinfo {author} {\bibfnamefont {L.}~\bibnamefont
  {Levitov}}, \ and\ \bibinfo {author} {\bibfnamefont {A.}~\bibnamefont
  {Chubukov}},\ }\href@noop {} {\bibfield  {journal} {\bibinfo  {journal}
  {Nature Physics}\ }\textbf {\bibinfo {volume} {8}},\ \bibinfo {pages} {158}
  (\bibinfo {year} {2012})}\BibitemShut {NoStop}%
\bibitem [{\citenamefont {Chichinadze}\ \emph
  {et~al.}(2020{\natexlab{a}})\citenamefont {Chichinadze}, \citenamefont
  {Classen},\ and\ \citenamefont {Chubukov}}]{PhysRevB.101.224513}%
  \BibitemOpen
  \bibfield  {author} {\bibinfo {author} {\bibfnamefont {D.~V.}\ \bibnamefont
  {Chichinadze}}, \bibinfo {author} {\bibfnamefont {L.}~\bibnamefont
  {Classen}}, \ and\ \bibinfo {author} {\bibfnamefont {A.~V.}\ \bibnamefont
  {Chubukov}},\ }\href {\doibase 10.1103/PhysRevB.101.224513} {\bibfield
  {journal} {\bibinfo  {journal} {Phys. Rev. B}\ }\textbf {\bibinfo {volume}
  {101}},\ \bibinfo {pages} {224513} (\bibinfo {year}
  {2020}{\natexlab{a}})}\BibitemShut {NoStop}%
\bibitem [{\citenamefont {Chichinadze}\ \emph
  {et~al.}(2020{\natexlab{b}})\citenamefont {Chichinadze}, \citenamefont
  {Classen},\ and\ \citenamefont {Chubukov}}]{PhysRevB.102.125120}%
  \BibitemOpen
  \bibfield  {author} {\bibinfo {author} {\bibfnamefont {D.~V.}\ \bibnamefont
  {Chichinadze}}, \bibinfo {author} {\bibfnamefont {L.}~\bibnamefont
  {Classen}}, \ and\ \bibinfo {author} {\bibfnamefont {A.~V.}\ \bibnamefont
  {Chubukov}},\ }\href {\doibase 10.1103/PhysRevB.102.125120} {\bibfield
  {journal} {\bibinfo  {journal} {Phys. Rev. B}\ }\textbf {\bibinfo {volume}
  {102}},\ \bibinfo {pages} {125120} (\bibinfo {year}
  {2020}{\natexlab{b}})}\BibitemShut {NoStop}%
\bibitem [{\citenamefont {Xu}\ and\ \citenamefont
  {Balents}(2018)}]{PhysRevLett.121.087001}%
  \BibitemOpen
  \bibfield  {author} {\bibinfo {author} {\bibfnamefont {C.}~\bibnamefont
  {Xu}}\ and\ \bibinfo {author} {\bibfnamefont {L.}~\bibnamefont {Balents}},\
  }\href {\doibase 10.1103/PhysRevLett.121.087001} {\bibfield  {journal}
  {\bibinfo  {journal} {Phys. Rev. Lett.}\ }\textbf {\bibinfo {volume} {121}},\
  \bibinfo {pages} {087001} (\bibinfo {year} {2018})}\BibitemShut {NoStop}%
\bibitem [{\citenamefont {Po}\ \emph {et~al.}(2018{\natexlab{b}})\citenamefont
  {Po}, \citenamefont {Zou}, \citenamefont {Vishwanath},\ and\ \citenamefont
  {Senthil}}]{PhysRevX.8.031089}%
  \BibitemOpen
  \bibfield  {author} {\bibinfo {author} {\bibfnamefont {H.~C.}\ \bibnamefont
  {Po}}, \bibinfo {author} {\bibfnamefont {L.}~\bibnamefont {Zou}}, \bibinfo
  {author} {\bibfnamefont {A.}~\bibnamefont {Vishwanath}}, \ and\ \bibinfo
  {author} {\bibfnamefont {T.}~\bibnamefont {Senthil}},\ }\href {\doibase
  10.1103/PhysRevX.8.031089} {\bibfield  {journal} {\bibinfo  {journal} {Phys.
  Rev. X}\ }\textbf {\bibinfo {volume} {8}},\ \bibinfo {pages} {031089}
  (\bibinfo {year} {2018}{\natexlab{b}})}\BibitemShut {NoStop}%
\bibitem [{\citenamefont {Yankowitz}\ \emph {et~al.}(2019)\citenamefont
  {Yankowitz}, \citenamefont {Chen}, \citenamefont {Polshyn}, \citenamefont
  {Zhang}, \citenamefont {Watanabe}, \citenamefont {Taniguchi}, \citenamefont
  {Graf}, \citenamefont {Young},\ and\ \citenamefont
  {Dean}}]{yankowitz2019tuning}%
  \BibitemOpen
  \bibfield  {author} {\bibinfo {author} {\bibfnamefont {M.}~\bibnamefont
  {Yankowitz}}, \bibinfo {author} {\bibfnamefont {S.}~\bibnamefont {Chen}},
  \bibinfo {author} {\bibfnamefont {H.}~\bibnamefont {Polshyn}}, \bibinfo
  {author} {\bibfnamefont {Y.}~\bibnamefont {Zhang}}, \bibinfo {author}
  {\bibfnamefont {K.}~\bibnamefont {Watanabe}}, \bibinfo {author}
  {\bibfnamefont {T.}~\bibnamefont {Taniguchi}}, \bibinfo {author}
  {\bibfnamefont {D.}~\bibnamefont {Graf}}, \bibinfo {author} {\bibfnamefont
  {A.~F.}\ \bibnamefont {Young}}, \ and\ \bibinfo {author} {\bibfnamefont
  {C.~R.}\ \bibnamefont {Dean}},\ }\href {\doibase 10.1126/science.aav1910}
  {\bibfield  {journal} {\bibinfo  {journal} {Science}\ }\textbf {\bibinfo
  {volume} {363}},\ \bibinfo {pages} {1059} (\bibinfo {year}
  {2019})}\BibitemShut {NoStop}%
\bibitem [{\citenamefont {Lu}\ \emph {et~al.}(2019)\citenamefont {Lu},
  \citenamefont {Stepanov}, \citenamefont {Yang}, \citenamefont {Xie},
  \citenamefont {Aamir}, \citenamefont {Das}, \citenamefont {Urgell},
  \citenamefont {Watanabe}, \citenamefont {Taniguchi}, \citenamefont {Zhang},
  \citenamefont {Bachtold}, \citenamefont {MacDonald},\ and\ \citenamefont
  {Efetov}}]{lu2019superconductors}%
  \BibitemOpen
  \bibfield  {author} {\bibinfo {author} {\bibfnamefont {X.}~\bibnamefont
  {Lu}}, \bibinfo {author} {\bibfnamefont {P.}~\bibnamefont {Stepanov}},
  \bibinfo {author} {\bibfnamefont {W.}~\bibnamefont {Yang}}, \bibinfo {author}
  {\bibfnamefont {M.}~\bibnamefont {Xie}}, \bibinfo {author} {\bibfnamefont
  {M.~A.}\ \bibnamefont {Aamir}}, \bibinfo {author} {\bibfnamefont
  {I.}~\bibnamefont {Das}}, \bibinfo {author} {\bibfnamefont {C.}~\bibnamefont
  {Urgell}}, \bibinfo {author} {\bibfnamefont {K.}~\bibnamefont {Watanabe}},
  \bibinfo {author} {\bibfnamefont {T.}~\bibnamefont {Taniguchi}}, \bibinfo
  {author} {\bibfnamefont {G.}~\bibnamefont {Zhang}}, \bibinfo {author}
  {\bibfnamefont {A.}~\bibnamefont {Bachtold}}, \bibinfo {author}
  {\bibfnamefont {A.~H.}\ \bibnamefont {MacDonald}}, \ and\ \bibinfo {author}
  {\bibfnamefont {D.~K.}\ \bibnamefont {Efetov}},\ }\href {\doibase
  10.1038/s41586-019-1695-0} {\bibfield  {journal} {\bibinfo  {journal}
  {Nature}\ }\textbf {\bibinfo {volume} {574}},\ \bibinfo {pages} {653}
  (\bibinfo {year} {2019})}\BibitemShut {NoStop}%
\bibitem [{\citenamefont {{Cao}}\ \emph {et~al.}(2020)\citenamefont {{Cao}},
  \citenamefont {{Rodan-Legrain}}, \citenamefont {{Park}}, \citenamefont {{Noah
  Yuan}}, \citenamefont {{Watanabe}}, \citenamefont {{Taniguchi}},
  \citenamefont {{Fernandes}}, \citenamefont {{Fu}},\ and\ \citenamefont
  {{Jarillo-Herrero}}}]{cao2020nematicity}%
  \BibitemOpen
  \bibfield  {author} {\bibinfo {author} {\bibfnamefont {Y.}~\bibnamefont
  {{Cao}}}, \bibinfo {author} {\bibfnamefont {D.}~\bibnamefont
  {{Rodan-Legrain}}}, \bibinfo {author} {\bibfnamefont {J.~M.}\ \bibnamefont
  {{Park}}}, \bibinfo {author} {\bibfnamefont {F.}~\bibnamefont {{Noah Yuan}}},
  \bibinfo {author} {\bibfnamefont {K.}~\bibnamefont {{Watanabe}}}, \bibinfo
  {author} {\bibfnamefont {T.}~\bibnamefont {{Taniguchi}}}, \bibinfo {author}
  {\bibfnamefont {R.~M.}\ \bibnamefont {{Fernandes}}}, \bibinfo {author}
  {\bibfnamefont {L.}~\bibnamefont {{Fu}}}, \ and\ \bibinfo {author}
  {\bibfnamefont {P.}~\bibnamefont {{Jarillo-Herrero}}},\ }\href@noop {}
  {\bibfield  {journal} {\bibinfo  {journal} {arXiv e-prints}\ ,\ \bibinfo
  {eid} {arXiv:2004.04148}} (\bibinfo {year} {2020})},\ \Eprint
  {http://arxiv.org/abs/2004.04148} {arXiv:2004.04148 [cond-mat.mes-hall]}
  \BibitemShut {NoStop}%
\bibitem [{\citenamefont {Jiang}\ \emph {et~al.}(2019)\citenamefont {Jiang},
  \citenamefont {Lai}, \citenamefont {Watanabe}, \citenamefont {Taniguchi},
  \citenamefont {Haule}, \citenamefont {Mao},\ and\ \citenamefont
  {Andrei}}]{jiang2019charge}%
  \BibitemOpen
  \bibfield  {author} {\bibinfo {author} {\bibfnamefont {Y.}~\bibnamefont
  {Jiang}}, \bibinfo {author} {\bibfnamefont {X.}~\bibnamefont {Lai}}, \bibinfo
  {author} {\bibfnamefont {K.}~\bibnamefont {Watanabe}}, \bibinfo {author}
  {\bibfnamefont {T.}~\bibnamefont {Taniguchi}}, \bibinfo {author}
  {\bibfnamefont {K.}~\bibnamefont {Haule}}, \bibinfo {author} {\bibfnamefont
  {J.}~\bibnamefont {Mao}}, \ and\ \bibinfo {author} {\bibfnamefont {E.~Y.}\
  \bibnamefont {Andrei}},\ }\href {\doibase 10.1038/s41586-019-1460-4}
  {\bibfield  {journal} {\bibinfo  {journal} {Nature}\ }\textbf {\bibinfo
  {volume} {573}},\ \bibinfo {pages} {91} (\bibinfo {year} {2019})}\BibitemShut
  {NoStop}%
\bibitem [{\citenamefont {Hejazi}\ \emph
  {et~al.}(2019{\natexlab{a}})\citenamefont {Hejazi}, \citenamefont {Liu},
  \citenamefont {Shapourian}, \citenamefont {Chen},\ and\ \citenamefont
  {Balents}}]{PhysRevB.99.035111}%
  \BibitemOpen
  \bibfield  {author} {\bibinfo {author} {\bibfnamefont {K.}~\bibnamefont
  {Hejazi}}, \bibinfo {author} {\bibfnamefont {C.}~\bibnamefont {Liu}},
  \bibinfo {author} {\bibfnamefont {H.}~\bibnamefont {Shapourian}}, \bibinfo
  {author} {\bibfnamefont {X.}~\bibnamefont {Chen}}, \ and\ \bibinfo {author}
  {\bibfnamefont {L.}~\bibnamefont {Balents}},\ }\href {\doibase
  10.1103/PhysRevB.99.035111} {\bibfield  {journal} {\bibinfo  {journal} {Phys.
  Rev. B}\ }\textbf {\bibinfo {volume} {99}},\ \bibinfo {pages} {035111}
  (\bibinfo {year} {2019}{\natexlab{a}})}\BibitemShut {NoStop}%
\bibitem [{\citenamefont {Hejazi}\ \emph
  {et~al.}(2019{\natexlab{b}})\citenamefont {Hejazi}, \citenamefont {Liu},\
  and\ \citenamefont {Balents}}]{PhysRevB.100.035115}%
  \BibitemOpen
  \bibfield  {author} {\bibinfo {author} {\bibfnamefont {K.}~\bibnamefont
  {Hejazi}}, \bibinfo {author} {\bibfnamefont {C.}~\bibnamefont {Liu}}, \ and\
  \bibinfo {author} {\bibfnamefont {L.}~\bibnamefont {Balents}},\ }\href
  {\doibase 10.1103/PhysRevB.100.035115} {\bibfield  {journal} {\bibinfo
  {journal} {Phys. Rev. B}\ }\textbf {\bibinfo {volume} {100}},\ \bibinfo
  {pages} {035115} (\bibinfo {year} {2019}{\natexlab{b}})}\BibitemShut
  {NoStop}%
\bibitem [{\citenamefont {Saito}\ \emph {et~al.}(2020)\citenamefont {Saito},
  \citenamefont {Ge}, \citenamefont {Watanabe}, \citenamefont {Taniguchi},\
  and\ \citenamefont {Young}}]{saito2020independent}%
  \BibitemOpen
  \bibfield  {author} {\bibinfo {author} {\bibfnamefont {Y.}~\bibnamefont
  {Saito}}, \bibinfo {author} {\bibfnamefont {J.}~\bibnamefont {Ge}}, \bibinfo
  {author} {\bibfnamefont {K.}~\bibnamefont {Watanabe}}, \bibinfo {author}
  {\bibfnamefont {T.}~\bibnamefont {Taniguchi}}, \ and\ \bibinfo {author}
  {\bibfnamefont {A.~F.}\ \bibnamefont {Young}},\ }\href {\doibase
  10.1038/s41567-020-0928-3} {\bibfield  {journal} {\bibinfo  {journal} {Nature
  Physics}\ }\textbf {\bibinfo {volume} {16}},\ \bibinfo {pages} {926}
  (\bibinfo {year} {2020})}\BibitemShut {NoStop}%
\bibitem [{\citenamefont {Xie}\ \emph {et~al.}(2019)\citenamefont {Xie},
  \citenamefont {Lian}, \citenamefont {J{\"a}ck}, \citenamefont {Liu},
  \citenamefont {Chiu}, \citenamefont {Watanabe}, \citenamefont {Taniguchi},
  \citenamefont {Bernevig},\ and\ \citenamefont
  {Yazdani}}]{xie2019spectroscopic}%
  \BibitemOpen
  \bibfield  {author} {\bibinfo {author} {\bibfnamefont {Y.}~\bibnamefont
  {Xie}}, \bibinfo {author} {\bibfnamefont {B.}~\bibnamefont {Lian}}, \bibinfo
  {author} {\bibfnamefont {B.}~\bibnamefont {J{\"a}ck}}, \bibinfo {author}
  {\bibfnamefont {X.}~\bibnamefont {Liu}}, \bibinfo {author} {\bibfnamefont
  {C.-L.}\ \bibnamefont {Chiu}}, \bibinfo {author} {\bibfnamefont
  {K.}~\bibnamefont {Watanabe}}, \bibinfo {author} {\bibfnamefont
  {T.}~\bibnamefont {Taniguchi}}, \bibinfo {author} {\bibfnamefont {B.~A.}\
  \bibnamefont {Bernevig}}, \ and\ \bibinfo {author} {\bibfnamefont
  {A.}~\bibnamefont {Yazdani}},\ }\href {\doibase 10.1038/s41586-019-1422-x}
  {\bibfield  {journal} {\bibinfo  {journal} {Nature}\ }\textbf {\bibinfo
  {volume} {572}},\ \bibinfo {pages} {101} (\bibinfo {year}
  {2019})}\BibitemShut {NoStop}%
\bibitem [{\citenamefont {{Tschirhart}}\ \emph {et~al.}(2020)\citenamefont
  {{Tschirhart}}, \citenamefont {{Serlin}}, \citenamefont {{Polshyn}},
  \citenamefont {{Shragai}}, \citenamefont {{Xia}}, \citenamefont {{Zhu}},
  \citenamefont {{Zhang}}, \citenamefont {{Watanabe}}, \citenamefont
  {{Taniguchi}}, \citenamefont {{Huber}},\ and\ \citenamefont
  {{Young}}}]{tschirhart2020imaging}%
  \BibitemOpen
  \bibfield  {author} {\bibinfo {author} {\bibfnamefont {C.~L.}\ \bibnamefont
  {{Tschirhart}}}, \bibinfo {author} {\bibfnamefont {M.}~\bibnamefont
  {{Serlin}}}, \bibinfo {author} {\bibfnamefont {H.}~\bibnamefont {{Polshyn}}},
  \bibinfo {author} {\bibfnamefont {A.}~\bibnamefont {{Shragai}}}, \bibinfo
  {author} {\bibfnamefont {Z.}~\bibnamefont {{Xia}}}, \bibinfo {author}
  {\bibfnamefont {J.}~\bibnamefont {{Zhu}}}, \bibinfo {author} {\bibfnamefont
  {Y.}~\bibnamefont {{Zhang}}}, \bibinfo {author} {\bibfnamefont
  {K.}~\bibnamefont {{Watanabe}}}, \bibinfo {author} {\bibfnamefont
  {T.}~\bibnamefont {{Taniguchi}}}, \bibinfo {author} {\bibfnamefont {M.~E.}\
  \bibnamefont {{Huber}}}, \ and\ \bibinfo {author} {\bibfnamefont {A.~F.}\
  \bibnamefont {{Young}}},\ }\href@noop {} {\bibfield  {journal} {\bibinfo
  {journal} {arXiv e-prints}\ ,\ \bibinfo {eid} {arXiv:2006.08053}} (\bibinfo
  {year} {2020})},\ \Eprint {http://arxiv.org/abs/2006.08053} {arXiv:2006.08053
  [cond-mat.mes-hall]} \BibitemShut {NoStop}%
\bibitem [{\citenamefont {Kang}\ and\ \citenamefont
  {Vafek}(2019)}]{PhysRevLett.122.246401}%
  \BibitemOpen
  \bibfield  {author} {\bibinfo {author} {\bibfnamefont {J.}~\bibnamefont
  {Kang}}\ and\ \bibinfo {author} {\bibfnamefont {O.}~\bibnamefont {Vafek}},\
  }\href {\doibase 10.1103/PhysRevLett.122.246401} {\bibfield  {journal}
  {\bibinfo  {journal} {Phys. Rev. Lett.}\ }\textbf {\bibinfo {volume} {122}},\
  \bibinfo {pages} {246401} (\bibinfo {year} {2019})}\BibitemShut {NoStop}%
\bibitem [{\citenamefont {Polshyn}\ \emph {et~al.}(2020)\citenamefont
  {Polshyn}, \citenamefont {Zhu}, \citenamefont {Kumar}, \citenamefont {Zhang},
  \citenamefont {Yang}, \citenamefont {Tschirhart}, \citenamefont {Serlin},
  \citenamefont {Watanabe}, \citenamefont {Taniguchi}, \citenamefont
  {MacDonald},\ and\ \citenamefont {Young}}]{polshyn2020electrical}%
  \BibitemOpen
  \bibfield  {author} {\bibinfo {author} {\bibfnamefont {H.}~\bibnamefont
  {Polshyn}}, \bibinfo {author} {\bibfnamefont {J.}~\bibnamefont {Zhu}},
  \bibinfo {author} {\bibfnamefont {M.~A.}\ \bibnamefont {Kumar}}, \bibinfo
  {author} {\bibfnamefont {Y.}~\bibnamefont {Zhang}}, \bibinfo {author}
  {\bibfnamefont {F.}~\bibnamefont {Yang}}, \bibinfo {author} {\bibfnamefont
  {C.~L.}\ \bibnamefont {Tschirhart}}, \bibinfo {author} {\bibfnamefont
  {M.}~\bibnamefont {Serlin}}, \bibinfo {author} {\bibfnamefont
  {K.}~\bibnamefont {Watanabe}}, \bibinfo {author} {\bibfnamefont
  {T.}~\bibnamefont {Taniguchi}}, \bibinfo {author} {\bibfnamefont {A.~H.}\
  \bibnamefont {MacDonald}}, \ and\ \bibinfo {author} {\bibfnamefont {A.~F.}\
  \bibnamefont {Young}},\ }\href {\doibase 10.1038/s41586-020-2963-8}
  {\bibfield  {journal} {\bibinfo  {journal} {Nature}\ }\textbf {\bibinfo
  {volume} {588}},\ \bibinfo {pages} {66} (\bibinfo {year} {2020})}\BibitemShut
  {NoStop}%
\bibitem [{\citenamefont {Bultinck}\ \emph {et~al.}(2020)\citenamefont
  {Bultinck}, \citenamefont {Chatterjee},\ and\ \citenamefont
  {Zaletel}}]{PhysRevLett.124.166601}%
  \BibitemOpen
  \bibfield  {author} {\bibinfo {author} {\bibfnamefont {N.}~\bibnamefont
  {Bultinck}}, \bibinfo {author} {\bibfnamefont {S.}~\bibnamefont
  {Chatterjee}}, \ and\ \bibinfo {author} {\bibfnamefont {M.~P.}\ \bibnamefont
  {Zaletel}},\ }\href {\doibase 10.1103/PhysRevLett.124.166601} {\bibfield
  {journal} {\bibinfo  {journal} {Phys. Rev. Lett.}\ }\textbf {\bibinfo
  {volume} {124}},\ \bibinfo {pages} {166601} (\bibinfo {year}
  {2020})}\BibitemShut {NoStop}%
\bibitem [{\citenamefont {Ochi}\ \emph {et~al.}(2018)\citenamefont {Ochi},
  \citenamefont {Koshino},\ and\ \citenamefont {Kuroki}}]{ochi2018possible}%
  \BibitemOpen
  \bibfield  {author} {\bibinfo {author} {\bibfnamefont {M.}~\bibnamefont
  {Ochi}}, \bibinfo {author} {\bibfnamefont {M.}~\bibnamefont {Koshino}}, \
  and\ \bibinfo {author} {\bibfnamefont {K.}~\bibnamefont {Kuroki}},\ }\href
  {\doibase 10.1103/PhysRevB.98.081102} {\bibfield  {journal} {\bibinfo
  {journal} {Phys. Rev. B}\ }\textbf {\bibinfo {volume} {98}},\ \bibinfo
  {pages} {081102} (\bibinfo {year} {2018})}\BibitemShut {NoStop}%
\bibitem [{\citenamefont {Chang}\ and\ \citenamefont
  {Niu}(2008)}]{chang2008berry}%
  \BibitemOpen
  \bibfield  {author} {\bibinfo {author} {\bibfnamefont {M.-C.}\ \bibnamefont
  {Chang}}\ and\ \bibinfo {author} {\bibfnamefont {Q.}~\bibnamefont {Niu}},\
  }\href@noop {} {\bibfield  {journal} {\bibinfo  {journal} {Journal of
  Physics: Condensed Matter}\ }\textbf {\bibinfo {volume} {20}},\ \bibinfo
  {pages} {193202} (\bibinfo {year} {2008})}\BibitemShut {NoStop}%
\bibitem [{\citenamefont {Xiao}\ \emph {et~al.}(2010)\citenamefont {Xiao},
  \citenamefont {Chang},\ and\ \citenamefont {Niu}}]{RevModPhys.82.1959}%
  \BibitemOpen
  \bibfield  {author} {\bibinfo {author} {\bibfnamefont {D.}~\bibnamefont
  {Xiao}}, \bibinfo {author} {\bibfnamefont {M.-C.}\ \bibnamefont {Chang}}, \
  and\ \bibinfo {author} {\bibfnamefont {Q.}~\bibnamefont {Niu}},\ }\href
  {\doibase 10.1103/RevModPhys.82.1959} {\bibfield  {journal} {\bibinfo
  {journal} {Rev. Mod. Phys.}\ }\textbf {\bibinfo {volume} {82}},\ \bibinfo
  {pages} {1959} (\bibinfo {year} {2010})}\BibitemShut {NoStop}%
\bibitem [{\citenamefont {He}\ \emph {et~al.}(2020)\citenamefont {He},
  \citenamefont {Goldhaber-Gordon},\ and\ \citenamefont {Law}}]{he2020giant}%
  \BibitemOpen
  \bibfield  {author} {\bibinfo {author} {\bibfnamefont {W.-Y.}\ \bibnamefont
  {He}}, \bibinfo {author} {\bibfnamefont {D.}~\bibnamefont
  {Goldhaber-Gordon}}, \ and\ \bibinfo {author} {\bibfnamefont {K.~T.}\
  \bibnamefont {Law}},\ }\href@noop {} {\bibfield  {journal} {\bibinfo
  {journal} {Nature Comm.}\ }\textbf {\bibinfo {volume} {11}},\ \bibinfo
  {pages} {1} (\bibinfo {year} {2020})}\BibitemShut {NoStop}%
\bibitem [{\citenamefont {{Huang}}\ \emph {et~al.}(2020)\citenamefont
  {{Huang}}, \citenamefont {{Wei}},\ and\ \citenamefont
  {{MacDoanld}}}]{MacDoanld2020}%
  \BibitemOpen
  \bibfield  {author} {\bibinfo {author} {\bibfnamefont {C.}~\bibnamefont
  {{Huang}}}, \bibinfo {author} {\bibfnamefont {N.}~\bibnamefont {{Wei}}}, \
  and\ \bibinfo {author} {\bibfnamefont {A.}~\bibnamefont {{MacDoanld}}},\
  }\href@noop {} {\bibfield  {journal} {\bibinfo  {journal} {arXiv e-prints}\
  ,\ \bibinfo {eid} {arXiv:2007.05990}} (\bibinfo {year} {2020})},\ \Eprint
  {http://arxiv.org/abs/2007.05990} {arXiv:2007.05990 [cond-mat.mes-hall]}
  \BibitemShut {NoStop}%
\bibitem [{\citenamefont {Kamenev}(2011)}]{kamenev2011field}%
  \BibitemOpen
  \bibfield  {author} {\bibinfo {author} {\bibfnamefont {A.}~\bibnamefont
  {Kamenev}},\ }\href@noop {} {\emph {\bibinfo {title} {Field theory of
  non-equilibrium systems}}}\ (\bibinfo  {publisher} {Cambridge University
  Press},\ \bibinfo {year} {2011})\BibitemShut {NoStop}%
\bibitem [{\citenamefont {Altland}\ and\ \citenamefont
  {Simons}(2010)}]{altland2010condensed}%
  \BibitemOpen
  \bibfield  {author} {\bibinfo {author} {\bibfnamefont {A.}~\bibnamefont
  {Altland}}\ and\ \bibinfo {author} {\bibfnamefont {B.~D.}\ \bibnamefont
  {Simons}},\ }\href@noop {} {\emph {\bibinfo {title} {Condensed matter field
  theory}}}\ (\bibinfo  {publisher} {Cambridge university press},\ \bibinfo
  {year} {2010})\BibitemShut {NoStop}%
\bibitem [{\citenamefont {Mermin}(1967)}]{mermin1967absence}%
  \BibitemOpen
  \bibfield  {author} {\bibinfo {author} {\bibfnamefont {N.~D.}\ \bibnamefont
  {Mermin}},\ }\href@noop {} {\bibfield  {journal} {\bibinfo  {journal}
  {Journal of Mathematical Physics}\ }\textbf {\bibinfo {volume} {8}},\
  \bibinfo {pages} {1061} (\bibinfo {year} {1967})}\BibitemShut {NoStop}%
\bibitem [{\citenamefont {Lifshitz}\ and\ \citenamefont
  {Pitaevskii}(1981)}]{LandauKinetics}%
  \BibitemOpen
  \bibfield  {author} {\bibinfo {author} {\bibfnamefont {E.}~\bibnamefont
  {Lifshitz}}\ and\ \bibinfo {author} {\bibfnamefont {L.~P.}\ \bibnamefont
  {Pitaevskii}},\ }\href@noop {} {\emph {\bibinfo {title} {Physical Kinetics,
  Volume 10 (Course of Theoretical Physics)}}}\ (\bibinfo  {publisher}
  {Pergamon Press, New York},\ \bibinfo {year} {1981})\BibitemShut {NoStop}%
\bibitem [{\citenamefont {Bistritzer}\ and\ \citenamefont
  {MacDonald}(2011)}]{bistritzer2011moire}%
  \BibitemOpen
  \bibfield  {author} {\bibinfo {author} {\bibfnamefont {R.}~\bibnamefont
  {Bistritzer}}\ and\ \bibinfo {author} {\bibfnamefont {A.~H.}\ \bibnamefont
  {MacDonald}},\ }\href@noop {} {\bibfield  {journal} {\bibinfo  {journal}
  {Proceedings of the National Academy of Sciences}\ }\textbf {\bibinfo
  {volume} {108}},\ \bibinfo {pages} {12233} (\bibinfo {year}
  {2011})}\BibitemShut {NoStop}%
\bibitem [{\citenamefont {Balents}(2019)}]{balents2019general}%
  \BibitemOpen
  \bibfield  {author} {\bibinfo {author} {\bibfnamefont {L.}~\bibnamefont
  {Balents}},\ }\href@noop {} {\bibfield  {journal} {\bibinfo  {journal}
  {SciPost Phys}\ }\textbf {\bibinfo {volume} {7}},\ \bibinfo {pages} {48}
  (\bibinfo {year} {2019})}\BibitemShut {NoStop}%
\bibitem [{\citenamefont {Kim}\ \emph {et~al.}(2018)\citenamefont {Kim},
  \citenamefont {Leconte}, \citenamefont {Chittari}, \citenamefont {Watanabe},
  \citenamefont {Taniguchi}, \citenamefont {MacDonald}, \citenamefont {Jung},\
  and\ \citenamefont {Jung}}]{kim2018accurate}%
  \BibitemOpen
  \bibfield  {author} {\bibinfo {author} {\bibfnamefont {H.}~\bibnamefont
  {Kim}}, \bibinfo {author} {\bibfnamefont {N.}~\bibnamefont {Leconte}},
  \bibinfo {author} {\bibfnamefont {B.~L.}\ \bibnamefont {Chittari}}, \bibinfo
  {author} {\bibfnamefont {K.}~\bibnamefont {Watanabe}}, \bibinfo {author}
  {\bibfnamefont {T.}~\bibnamefont {Taniguchi}}, \bibinfo {author}
  {\bibfnamefont {A.~H.}\ \bibnamefont {MacDonald}}, \bibinfo {author}
  {\bibfnamefont {J.}~\bibnamefont {Jung}}, \ and\ \bibinfo {author}
  {\bibfnamefont {S.}~\bibnamefont {Jung}},\ }\href {\doibase
  10.1021/acs.nanolett.8b03423} {\bibfield  {journal} {\bibinfo  {journal}
  {Nano Lett.}\ }\textbf {\bibinfo {volume} {18}},\ \bibinfo {pages} {7732}
  (\bibinfo {year} {2018})}\BibitemShut {NoStop}%
\end{thebibliography}%

\clearpage

\appendix
\onecolumngrid
                            %                             \begin{widetext}

\section{Valley Polarization Order Parameter Dynamics from Keldysh Formalism}
\label{Sec:VPOPDynamicsFromKeldysh}

This section is for the demonstration of the valley polarization in the twisted bilayer graphene (tBLG) with the presence of an external bias. The `slow' dynamics of the valley polarization order parameter, denoted by $\Phi_{\text{cl}}(\boldsymbol{x},t)$ and its Fourier transformation $\Phi_{\text{cl}}(\boldsymbol{q},\omega)$, is governed by the following action:
\begin{equation}
  \begin{split}
    \mathcal{F}[\Phi_{\text{cl}},\Phi_{\text{q}}]=&U\int\frac{d\omega}{2\pi}\int\frac{d^2q}{(2\pi)^2}\ \alpha_2(\boldsymbol{q},\omega)\Phi_{\text{q}}(-\boldsymbol{q},-\omega)\Phi_{\text{cl}}(\boldsymbol{q},\omega)\\
    +&U\int_{-\infty}^{\infty}dt\int d^2x\left[\alpha_1(\boldsymbol{x},t)\Phi_{\text{q}}(\boldsymbol{x},t)+\alpha_3(\boldsymbol{x},t)\Phi_{\text{q}}(\boldsymbol{x},t)\Phi^2_{\text{cl}}(\boldsymbol{x},t)+\alpha_4(\boldsymbol{x},t)\Phi_{\text{q}}(\boldsymbol{x},t)\Phi^3_{\text{cl}}(\boldsymbol{x},t)\right]\\
    +&\mathcal{O}\left(\Phi_{\text{q}}^2\right)
  \end{split}
\end{equation}
with the saddle-point time evolution equation given by:
\begin{equation}
  0=\left.\frac{\delta\mathcal{F}[\Phi_{\text{cl}},\Phi_{\text{q}}]}{\delta \Phi_{\text{q}}}\right|_{\Phi_{\text{q}}=0}
\end{equation}
Here, $\Phi_{\text{q}}$ corresponds to the quantum fluctuations of the valley polarization, in the Keldysh language.

The coefficients of $\alpha_i$ is listed below:
\begin{eqnarray}
  &\alpha_1(\boldsymbol{x},t)&=n^{(1)}(\boldsymbol{x},t)-n^{(2)}(\boldsymbol{x},t)=\Delta n_0(\boldsymbol{x},t)\\
  &\alpha_2(\boldsymbol{q},\omega)&=U(\nu+\frac{\pi^2}{3}\nu^{\prime\prime}T^2+\nu\frac{i\omega}{Dq^2-i\omega})-1\\
  &\alpha_3(\boldsymbol{x},t)&=-\frac{1}{8}U^2\int\frac{d^2k}{(2\pi)^2}\partial^2_{\epsilon_{\boldsymbol{k}}}\left[f^{(1)}(\boldsymbol{k})-f^{(2)}(\boldsymbol{k})\right]\\
  &\alpha_4(\boldsymbol{x},t)&=\frac{1}{48}U^3\int\frac{d^2k}{(2\pi)^2}\partial^3_{\epsilon_{\boldsymbol{k}}}\left[f^{(1)}(\boldsymbol{k})+f^{(2)}(\boldsymbol{k})\right]=\frac{1}{24}U^3\nu^{\prime\prime}
\end{eqnarray}

Notations:
\begin{enumerate}
\item $\Phi_{\text{cl}}$ corresponds to the VPOP $\Phi_{\text{v}}$ in the maintext;
\item $n^{(i)}(\boldsymbol{x},t)$ is the electron density of valley $(i)$;
\item $\nu=\nu(\epsilon_{\text{F}})$ is the electron density of states at Fermi level $\epsilon_{\text{F}}$ of a given valley, while $\nu^{\prime\prime}$ is the second derivative of density of states. The valley polarized state is stable when $\nu^{\prime\prime}<0$.
\item $U$ is the Stoner interaction strength;
\item $D\approx\frac{1}{2}v_{\text{F}}^2\tau$ is the electron's diffusion constant;
\item $T$ is the temperature;
\item $\alpha_3$ will be explained in detail later. It involves the difference of the electron's distribution function in the two valleys, $f^{i}(\epsilon,\theta)$. Thus, it is proportional to the bias.
\end{enumerate}

\subsection{Model Hamiltonian}

We consider a model of the following Hamiltonian:
\begin{equation}
  \hat{H}=\hat{H}_0+V_{\text{dis}}+\hat{H}_{\text{int}}
\end{equation}

The first part of the Hamiltonian is given by:
\begin{equation}
  \hat{H}_0=\begin{bmatrix}
    H^{(+)}(-i\partial_{\boldsymbol{x}})+V_{\text{bias}}(\boldsymbol{x}) & 0\\
    0 & H^{(-)}(-i\partial_{\boldsymbol{x}})+V_{\text{bias}}(\boldsymbol{x})
  \end{bmatrix}
\end{equation}
Electrons live in the two valleys described by $H^{(s)}(-i\partial_{\boldsymbol{x}})$. The two valleys are presumed to be time reversal (TR) related:
\begin{equation}
  \mc{T}\,H^{(+)}(-i\partial_{\boldsymbol{x}})\,\mc{T}^{-1}=H^{(-)}(i\partial_{\boldsymbol{x}})
\end{equation}
The system is subject to a bias electric potential $V_{\text{bias}}(\boldsymbol{x})$. 

The second part is the disorder potential:
\begin{equation}
  V_{\text{dis}}=\begin{bmatrix}
    V_0(\boldsymbol{x}) & V_1(\boldsymbol{x})\\
    V_1(\boldsymbol{x}) & V_0(\boldsymbol{x})
  \end{bmatrix}
  \label{appeq:disorder}
\end{equation}
The electrons experiences intra-valley impurity scattering potential $V_0(\boldsymbol{x})$ and the inter-valley impurity scattering potential $V_1(\boldsymbol{x})$. The impurity potentials follow the following probability distribution:
\begin{equation}
  \begin{split}
    &P[V_0(\boldsymbol{x})]=\text{Exp}\left[-\pi\nu\tau\int d^2x\left|V_0(\boldsymbol{x})\right|^2\right]\\
    &P[V_1(\boldsymbol{x})]=\text{Exp}\left[-\pi\nu\tau^{\prime}\int d^2x\left|V_1(\boldsymbol{x})\right|^2\right]
  \end{split}
\end{equation}
and following correlation:
\begin{equation}
  \langle V_0(\boldsymbol{x})V_0(\boldsymbol{x}^{\prime})\rangle=\frac{\delta^{(2)}(\boldsymbol{x}-\boldsymbol{x}^{\prime})}{2\pi\nu\tau};\ \ \ \ \ \ \ \ 
  \langle V_1(\boldsymbol{x})V_1(\boldsymbol{x}^{\prime})\rangle=\frac{\delta^{(2)}(\boldsymbol{x}-\boldsymbol{x}^{\prime})}{2\pi\nu\tau^{\prime}}
\end{equation}
where $\langle\cdots\rangle$ means disorder average.

The third part of the Hamiltonian gives the Stoner interaction between two valleys:
\begin{equation}
  \hat{H}_{\text{int}}=Un^{(+)}(\boldsymbol{x},t)n^{(-)}(\boldsymbol{x},t)
\end{equation}
If the interaction is strong enough to the system may develop valley polarization spontaneously at low temperature. 

\subsection{Keldysh Formulation}

For nonequilibrium and disordered system, it's convenient to use Keldysh formulation to extract the physical features. The formulation is based on the following path integral:
\begin{equation}
  \langle\mathcal{Z}\rangle=\int D\bar{\psi}^{(+)}D\psi^{(+)}D\bar{\psi}^{(-)}D\psi^{(-)} e^{i\mathcal{S}[\bar{\psi}^{(+)},\psi^{(+)},\bar{\psi}^{(-)},\psi^{(-)}]}
\end{equation}
with the action given by:
\begin{equation}
  \mathcal{S}=\int_{\mathcal{C}}dt\int d^2x\left\{[\bar{\psi}^{(+)},\bar{\psi}^{(-)}]\left[i\partial_t-\hat{H}_0\right]\begin{bmatrix}
      \psi^{(+)}\\
      \psi^{(-)}
    \end{bmatrix}-U\bar{\psi}^{(+)}\psi^{(+)}\bar{\psi}^{(-)}\psi^{(-)}\right\}
\end{equation}
The time contour is defined as $\mathcal{C}=\left\{-\infty,\infty\right\}\cup\left\{\infty,-\infty\right\}$, going from negative infinity to infinity then back to negative infinity.

The dynamics of the valley polarization order parameter may be obtained by a Hubbard-Strantonovich transformation:
\begin{equation}
  \langle\mathcal{Z}\rangle=\langle\int D\bar{\psi}D\psi D\Phi e^{i\mathcal{S}[\bar{\psi},\psi,\Phi]}\rangle
\end{equation}
with the new action:
\begin{equation}
  \mathcal{S}[\bar{\psi},\psi,h]=\int_{\mathcal{C}}dt\int d^2x\left\{\bar{\psi}\left[i\partial_t-\hat{H}_0\right]
    \psi-\frac{1}{4}U\left[\bar{\psi}\psi\right]^2+\frac{1}{2}U\Phi\bar{\psi}\sigma_z\psi-\frac{1}{4}U\Phi^2\right\}
\end{equation}
Here, the fermionic degrees of freedom is compactly written as $\bar{\psi}=[\bar{\psi}^{(+)},\bar{\psi}^{(-)}]$ and $\psi=[\psi^{(+)},\psi^{(-)}]^{\text{T}}$. $\sigma_z$ is the Pauli matrix in the valley space. The valley polarization order parameter $\Phi$ couples to the difference of the electron densities in the two valleys $\bar{\psi}\sigma_z\psi$, and the saddle point solution reads:
\begin{equation}
\Phi=\langle \bar{\psi}\sigma_z \psi \rangle
\end{equation}

It's convenient to perform a Keldysh rotation before proceeding further:
\begin{equation}
  \bar{\psi}_{1/2}=\frac{1}{\sqrt{2}}\left(\bar{\psi}_{+}\mp\bar{\psi}_{-}\right),\ \ \ \psi_{1/2}=\frac{1}{\sqrt{2}}\left(\psi_{+}\pm\psi_{-}\right),\ \ \ \Phi_{\text{cl/q}}=\frac{1}{2}\left(\Phi_{+}\pm \Phi_{-}\right)
\end{equation}
Here, the subindex $+(-)$ indicates the fields on the forward (backward) time domain $\left\{-\infty,\infty\right\}$ ($\left\{\infty,-\infty\right\}$).

After the Keldysh rotation, the action reads:
\begin{equation}
  \mathcal{S}=\int_{-\infty}^{\infty}dt\int d^2x\left\{\check{\bar{\psi}}\left[G_0^{-1}-V_{\text{dis}}\gamma^{\text{cl}}+\frac{1}{2}U\Phi_{\alpha}\sigma_z\gamma^{\alpha}\right]\check{\psi}-U\Phi_{\text{cl}}\Phi_{\text{q}}\right\}
\end{equation}
Here, we assume the total density $\bar{\psi}\psi$ is fixed, thus neglected the term of $-U\left[\bar{\psi}\psi\right]^2$. The meaning of $\Phi_{\alpha}\gamma^{\alpha}$ is explained below.

Some notations: The fermionic fields are two component spinor in Keldysh space, $\check{\bar{\psi}}=[\bar{\psi}_1,\bar{\psi}_2]$ and $\check{\psi}=[\psi_1,\psi_2]^{\text{T}}$. And each component is also a two component spinor in valley space, $\psi_{1/2}=[\psi^{(+)}_{1/2},\psi^{(-)}_{1/2}]^{\text{T}}$ and similarly for $\bar{\psi}_{1/2}$. $V_{\text{dis}}$ is the disorder potential and is a two by two matrix in the valley space.  Meanwhile, $\gamma^{\alpha}$ with $\alpha=\text{cl, q}$ are the matrices in the Keldysh space:
\begin{equation}
  \gamma^{\text{cl}}=\begin{bmatrix}
    1&0\\0&1
  \end{bmatrix};\ \gamma^{\text{q}}=\begin{bmatrix}
    0&1\\
    1&0
  \end{bmatrix}.
\end{equation}

Then, the fermionic degrees of freedom may be integrated out directly:
\begin{equation}
  \langle\mathcal{Z}\rangle=\langle\int Dh\ \exp\left\{-i U\int_{-\infty}^{\infty}dt\int d^2x\Phi_{\text{cl}}\Phi_{\text{q}}+\tr\ln \left[G^{-1}_0-V_{\text{dis}}\gamma^{\text{cl}}+\frac{U}{2}\Phi_{\alpha}\sigma_z\gamma^{\alpha}\right]\right\}\rangle
  \label{appeq:Z}
\end{equation}
where $\tr\ln[...]$ includes summation over space, time, Keldysh and internal valley d.o.f. This is a path integral with an action depends on the order parameter. The goal is to find the effective disorder averaged action $\mathcal{F}[\Phi_{\text{cl}},\Phi_{\text{q}}]=-i \ln \langle \mc{Z} \rangle$ that is linear in $\Phi_{\text{q}}$, so that that the semiclassical dynamics of the order parameter is given by:
\begin{equation}
  0=\left.\frac{\delta\mathcal{F}[\Phi_{\text{cl}},\Phi_{\text{q}}]}{\delta \Phi_{\text{q}}}\right|_{\Phi_{\text{q}}=0}
\end{equation}

\subsection{Disorder Averaging Process}

The first observation is that $\mathcal{S}[\Phi_{\text{cl}},\Phi_{\text{q}}=0]=0$. A direct implication is that when we do power expansion of the $\tr\ln\left[\cdots\right]$ in powers of the order parameter $\Phi_{\alpha}$, each term is at least linear in $\Phi_{\text{q}}$. Thus, the expansion goes like follows:
\begin{align}
  \tr\ln \left[G^{-1}_0-V_{\text{dis}}\gamma^{\text{cl}}+\frac{1}{2}U\Phi_{\alpha}\sigma_z\gamma^{\alpha}\right]&=\sum_{n=1}^{\infty}\frac{(-1)^{n-1}}{n}\tr\left[\left(G_0^{-1}-V_{\text{dis}}\gamma^{\text{cl}}\right)^{-1}\frac{1}{2}U\Phi_{\alpha}\sigma_z\gamma^{\alpha}\right]^n\non\\
                                                                                                             &=\sum_{n=1}^{\infty}\frac{(-1)^{n-1}}{n}\tr\left[\Gb\frac{1}{2}U\Phi_{\alpha}\sigma_z\gamma^{\alpha}\right]^n
\end{align}
where for brevity, we define $\Gb=\left(G_0^{-1}-V_{\text{dis}}\gamma^{\text{cl}}\right)^{-1}$. 
Then, we expand the exponential and then do the disorder average and then re-exponentiate the expression. During this process, we keep our accuracy only to linear order in $\Phi_{\text{q}}$.
\begin{enumerate}
\item Expand the exponential to linear order in $\Phi_{\text{q}}$:
  \begin{equation}
    \exp\left\{\tr\ln\left[\cdots\right]\right\}=1+\sum_{n=1}^{\infty}\frac{(-1)^{n-1}}{n}\tr\left[\Gb\frac{1}{2}U\Phi_{\alpha}\sigma_z\gamma^{\alpha}\right]^n+\mathcal{O}\left(\Phi_{\text{q}}^2\right),
  \end{equation}
\item Do the disorder average and keeping terms up to $\Phi^4$:
  \begin{equation}
    \begin{split}
      \langle\exp\left\{\tr\ln\left[\cdots\right]\right\}\rangle
      =&1+U\langle\tr\left[\Gb\frac{1}{2}\Phi_{\alpha}\sigma_z\gamma^{\alpha}\right]\rangle
      -\frac{1}{2}U^2\tr\langle\left[\Gb\frac{1}{2}\Phi_{\alpha}\sigma_z\gamma^{\alpha}\right]^2\rangle\\
      &+\frac{1}{3}U^3\tr\langle\left[\Gb\frac{1}{2}\Phi_{\alpha}\sigma_z\gamma^{\alpha}\right]^3\rangle
      -\frac{1}{4}U^4\tr\langle\left[\Gb\frac{1}{2}\Phi_{\alpha}\sigma_z\gamma^{\alpha}\right]^4\rangle+\cdots
    \end{split}
    \label{appeq:A25}
  \end{equation}
\item After re-exponentiate Eq.~\eqref{appeq:A25} and including the non-interacting quadratic in $\Phi$ term from Eq.~\eqref{appeq:Z}, we find the effective disorder averaged action as 
  \begin{align}
    \mc{F}=&-iU\tr[\Phi_{\text{cl}}\Phi_{\text{q}}]+U\langle\tr\left[\Gb\frac{1}{2}\Phi_{\alpha}\sigma_z\gamma^{\alpha}\right]\rangle-\frac{1}{2}U^2\tr\langle\left[\Gb\frac{1}{2}\Phi_{\alpha}\sigma_z\gamma^{\alpha}\right]^2\rangle\non\\
           &+\frac{1}{3}U^3\tr\langle\left[\Gb\frac{1}{2}\Phi_{\alpha}\sigma_z\gamma^{\alpha}\right]^3\rangle-\frac{1}{4}U^4\tr\langle\left[\Gb\frac{1}{2}\Phi_{\alpha}\sigma_z\gamma^{\alpha}\right]^4\rangle+\cdots
  \end{align}
            
\end{enumerate}
Notice that this process of disorder averaging is quite straightforward here. This is because we aim at the semi-classical dynamics of the order parameter and keep our accuracy only to linear order in $\Phi_{\text{q}}$. Thus, different terms do not mix (since each term is already linear in $\Phi_{\text{q}}$.)

\subsection{The meaning of each term}

\begin{enumerate}
\item The linear term vanishes in equilibrium due to time reversal symmetry. A bias electric potential may lead to non-zero value as we show in Sec.~\ref{appsec:A5}. % $\sim U\frac{\Phi_{\text{q}}}{2}\langle\tr\left[\Gb\sigma_z\gamma^{q}\right]\rangle$ contains the disorder averaged Green's function $G=\langle\Gb\rangle$;
\item The quadratic term contains the polarization operator:
  \begin{equation}
    \sim-\frac{U^2}{2}\Phi_{\text{q}}\Phi_{\text{cl}}\tr\langle\left[\Gb\sigma_z\gamma^{\text{cl}}\Gb\sigma_z\gamma^{\text{q}}\right]^2\rangle= iU^2\Phi_{\text{q}}\Phi_{\text{cl}}\Pi(q,\omega)
  \end{equation}
  % The polarization operator $\Pi(q,\omega)$ might take RPA or diffuson value, depending on the frequency and momentum; 
  The polarization $\Pi(q,\omega)$ in the static limit contributes to the susceptibility of valley polarization order parameter, and drives a 2nd order phase transition to valley polarized state below $T_c$.
\item Similar to the linear term, the cubic term vanishes in equilibrium due to time reversal symmetry as well. To analyze the leading order non-equilibrium effect due to bias potential, we keep the linear term only and ignore the cubic term, which is smaller by $\Phi_{\text{cl}}^2$ near $T_c$. 
\item The quartic term should proportional to the second derivative of the electron's density of states as in the usual description of the Stoner instability.
\end{enumerate}
To summarize, the non-equilibrium effect is mainly captured by the linear term. The rest captures the interaction effect to the effective action in equilibrium, which has been studied well in the context of Stoner instability. As a result, in the following perturbative expansion in $\Phi$, $V_{\text{bias}}$ is considered only in the linear term. To obtain the coefficients for quadratic and quartic terms, we consider $V_{\text{bias}}=0$.

\subsection{Linear Term}
\label{appsec:A5}
The linear term we are chasing after only contains the Keldysh Green's function:
\begin{equation}
  \langle \tr\left[\Gb\sigma_z\gamma^{q}\frac{\Phi_{\text{q}}}{2}\right]\rangle=\int_{-\infty}^{\infty}dt\int d^2x\frac{\Phi_{\text{q}}(\boldsymbol{x},t)}{2}\tr\left[\langle\Gb^{\text{K}}(\boldsymbol{x},\boldsymbol{x};t,t)\rangle\sigma_z\right]
\end{equation}

The equal spacetime Keldysh Green's function is the distribution function up to gradient corrections:
\begin{equation}
  \langle\Gb^{\text{K}}\rangle(\boldsymbol{x},\boldsymbol{x};t,t)=\int\frac{d\omega}{2\pi}\int\frac{d^2k}{(2\pi)^2}F(\boldsymbol{x},t;\boldsymbol{k},\omega)\left[\langle\Gb^{\text{R}}\rangle(\boldsymbol{x},t;\boldsymbol{k},\omega)-\langle\Gb^{\text{A}}\rangle(\boldsymbol{x},t;\boldsymbol{k},\omega)\right]
\end{equation}
The difference if retarded and advanced Green's function is a delta function $G^{\text{R}}-G^{\text{A}}=-i2\pi\delta\left(\omega-H_0\right)$. The integration over the frequency puts $F$ on-mass shell, making a real distribution function. Then, integration over momentum gives $\sim 1-2n(\boldsymbol{x},t)$, with $n(\boldsymbol{x},t)$ being the electron density. Note that $F$ is traced with $\sigma_z$. Thus,
\begin{equation}
  \boxed{U\langle \tr\left[\Gb\sigma_z\gamma^{q}\frac{\Phi_{\text{q}}}{2}\right]\rangle=iU\int_{-\infty}^{\infty}dt\int d^2x\ \Phi_{\text{q}}(\boldsymbol{x},t)\ \left[n^{(1)}(\boldsymbol{x},t)-n^{(2)}(\boldsymbol{x},t)\right]}
\end{equation}
the order parameter $\Phi_{\text{q}}$ couples to the difference in the electron density of the two valleys. 

The valley density difference is defined as $\Delta n_0(\boldsymbol{x},t)=n^{(1)}(\boldsymbol{x},t)-n^{(2)}(\boldsymbol{x},t) $ hereafter. $\Delta n_0(\boldsymbol{x},t)$ is induced by the bias field only when proper inter-valley scattering is taken into account. For simplicity, we will ignore the electron interaction to obtain $\Delta n_0$. Formally, the self-consistent kinetic equation for $F$ and thus $\Delta n_0$ can be obtained as below. 
% As discussed, the linear term contains the disorder averaged Green's function. 

In Keldysh space, the fermionic Green's function has the following structure:
\begin{equation}
  \Gb=\begin{bmatrix}
    \Gb^{\text{R}}&\Gb^{\text{K}}\\
    0&\Gb^{\text{A}}
  \end{bmatrix}
\end{equation}

The Green's function is a function of two space-time coordinates, $\Gb=\Gb(\boldsymbol{x},\boldsymbol{x}^{\prime};t,t^{\prime})$. It can be written in terms of Wigner coordinates:
\begin{equation}
  \Gb(\boldsymbol{x},t;\boldsymbol{k},\omega)=\int d\Delta t\int d^2\Delta x \ e^{-i\boldsymbol{k}\cdot\Delta\boldsymbol{x}+i\omega\Delta t}\Gb(\boldsymbol{x}+\frac{\Delta\boldsymbol{x}}{2},\boldsymbol{x}-\frac{\Delta\boldsymbol{x}}{2};t+\frac{\Delta t}{2},t-\frac{\Delta t}{2})
\end{equation}

The retarded and advanced Green's functions are given by the standard disorder calculation:
\begin{equation}
  G_0^{\text{R/A}}(\boldsymbol{x},t;\boldsymbol{k},\omega)=\frac{1}{\omega-H_0(\boldsymbol{k},\boldsymbol{x})\pm i\delta}+\text{Gradient Corrections}
\end{equation}

The Keldysh Green's function may be parameterized as $G^{\text{K}}=G^{\text{R}}\star F-F\star G^{\text{A}}$ (the Wigner coordinates are not written explicitly). The star operation is defined as: 
\begin{equation}
  \star=\exp\left\{\frac{i}{2}\left[\overleftarrow{\partial}_{\boldsymbol{x}}\cdot\overleftrightarrow{\partial}_{\boldsymbol{k}}-\overleftarrow{\partial}_{\boldsymbol{k}}\cdot\overleftrightarrow{\partial}_{\boldsymbol{x}}-\overleftarrow{\partial}_{t}\cdot\overleftrightarrow{\partial}_{\omega}+\overleftarrow{\partial}_{\omega}\cdot\overleftrightarrow{\partial}_{t}\right]\right\}
\end{equation}
$F$ plays the role of density matrix with a two by two structure in valley space, satisfying the following equation:
\begin{equation}
  -\left[\omega-H_0\stackrel{\star}{,}F\right]=\Sigma^{\text{K}}-\left(\Sigma^{\text{R}}\star F-F\star\Sigma^{\text{A}}\right)
\end{equation}
This formal equation is essentially the Boltzmann equation in some simple cases (neglecting the entanglement between two valleys). It needs to be solved independently. Within mass shell approximation of $F$, and considering impurity scattering of the form Eq.~\eqref{appeq:disorder}, we obtain the semi-classical Boltzmann equation [Eq.~\eqref{Eq:SBE0}] in the main text. While the band carries non-zero Chern number, we have checked that the Berry curvature effect does not contribute to valley polarization in the linear response, so it is ignored to obtain Eq.~\eqref{Eq:SBE0}.

\subsection{The quadratic term}

For the quadratic term, only the equilibrium contribution needs to be considered for our purpose. The quadratic term from interaction reads:
\begin{align}
  &-\frac{1}{2}U^2\tr\langle\left[\Gb\frac{1}{2}\Phi_{\alpha}\sigma_z\gamma^{\alpha}\right]^2\rangle\non\\
  =&-\frac{1}{8}U^2\sum_{\alpha,\beta}\int_{-\infty}^{\infty}dt_1dt_2\int d^2x_1d^2x_2 \Phi_{\alpha}(\boldsymbol{x}_2,t_2)\Phi_{\beta}(\boldsymbol{x}_1,t_1)\tr\langle G_{\text{b}}(\boldsymbol{x}_1,\boldsymbol{x}_2;t_1,t_2)\sigma_z\gamma^{\alpha}G_{\text{b}}(\boldsymbol{x}_2,\boldsymbol{x}_1;t_2,t_1)\sigma_z\gamma^{\beta}\rangle
\end{align}
     %      where $G_{\text{b}}=\left(G_0^{-1}-V_{\text{dis}}\gamma^{\text{cl}}\right)^{-1}$ is the bare Green's function.

Some observations:
\begin{enumerate}
\item Only the term of the form $\Phi_{\text{q}}\Phi_{\text{cl}}$ is relevant. The associate coefficient is $\sim\tr\left[ \Gb\sigma_z\gamma^{\text{cl}}\Gb\sigma_z\gamma^{\text{q}}\right]\sim\tr\left[\Gb^{\text{R}}\Gb^{\text{K}}+\Gb^{\text{K}}\Gb^{\text{A}}\right]$;
\item Current treatment does not have explicit time dependence in the Hamiltonian. Thus, the Green's functions are functions of time difference;
\item At equilibrium, the Keldysh Green's function is of the following form:
  \begin{equation}
    \langle\Gb^{\text{K}}\rangle(\boldsymbol{x}_1,\boldsymbol{x}_2;t_1,t_2)=\int\frac{d\epsilon}{2\pi}e^{-i\epsilon\left(t_1-t_2\right)}F(\epsilon)\left[\langle\Gb^{\text{R}}\rangle(\boldsymbol{x}_1,\boldsymbol{x}_2;\epsilon)-\langle\Gb^{\text{A}}\rangle(\boldsymbol{x}_1,\boldsymbol{x}_2;\epsilon)\right];
  \end{equation}
\item For simplicity, the inter-valley scattering in $\Gb$ is ignored. As a result, the Green's functions are diagonal in the valley space. We will argue below that the simplification only modify the result quantitatively. 
  % As we show below, the disorder scattering leads to diffuson collective mode in $\Pi$, which is generic as a result of replica symmetry breaking, inter-valley scattering should only modify the diffusion constant quantitatively. %This assumption will simplify the calculations later. 
\end{enumerate}

More careful analysis shows that the quadratic term is:
\begin{equation}
  \begin{split}
    &-\frac{1}{2}U^2\tr\langle\left[\Gb\frac{1}{2}\Phi_{\alpha}\sigma_z\gamma^{\alpha}\right]^2\rangle\\
    =&-\frac{1}{4}U^2\int_{-\infty}^{\infty}dt_1dt_2\int d^2x_1d^2x_2 \Phi_{\text{q}}(\boldsymbol{x}_2,t_2)\Phi_{\text{cl}}(\boldsymbol{x}_1,t_1)\\
    \times&\tr\langle \Gb^{\text{K}}(\boldsymbol{x}_1,\boldsymbol{x}_2;t_1,t_2)\sigma_z \Gb^{\text{R}}(\boldsymbol{x}_2,\boldsymbol{x}_1;t_2,t_1)\sigma_z+\Gb^{\text{A}}(\boldsymbol{x}_1,\boldsymbol{x}_2;t_1,t_2)\sigma_z \Gb^{\text{K}}(\boldsymbol{x}_2,\boldsymbol{x}_1;t_2,t_1)\sigma_z\rangle
  \end{split}
\end{equation}
What's in the trace should be $G^{\text{R}}\sigma_zG^{\text{K}}\sigma_z+G^{\text{K}}\sigma_zG^{\text{A}}\sigma_z$. Since we assumed the Green's functions are diagonal in the valley space, the summation over valley index only contribute to a factor 2.
% Such assumption also assumes the impurity potential to be diagonal. This is equivalently assuming that the intra-valley impurity scattering is much stronger than that of inter-valley one. 

Next step is to rewrite the fields and Green's functions in frequency space, one arrives at the following expression:
\begin{equation}
  \begin{split}
    &-\frac{U^2}{2}\int\frac{d\omega}{2\pi}\int d^2x_1d^2x_2\Phi_{\text{q}}(\boldsymbol{x}_2,-\omega)\Phi_{\text{cl}}(\boldsymbol{x}_1,\omega)\\
    \times&\int\frac{d\epsilon}{2\pi}\langle \Gb^{\text{K}}(\boldsymbol{x}_1,\boldsymbol{x}_2;\epsilon)\Gb^{\text{R}}(\boldsymbol{x}_2,\boldsymbol{x}_1;\epsilon+\omega)+\Gb^{\text{A}}(\boldsymbol{x}_1,\boldsymbol{x}_2;\epsilon)\Gb^{\text{K}}(\boldsymbol{x}_2,\boldsymbol{x}_1;\epsilon+\omega)\rangle
  \end{split}
\end{equation}
where the valley d.o.f. has been summed over. 
The second line can be further expressed as
            %             The second line is the usual polarization operator. This can be seen by rewriting the Keldysh Green's function in terms of retarded/advanced ones and the distribution function:
\begin{align}
  &\int\frac{d\epsilon}{2\pi}\left[\langle \Gb^{\text{R}}(\boldsymbol{x}_1,\boldsymbol{x}_2;\epsilon+\omega) \Gb^{\text{R}}(\boldsymbol{x}_2,\boldsymbol{x}_1;\epsilon)F(\epsilon+\omega)-\Gb^{\text{A}}(\boldsymbol{x}_1,\boldsymbol{x}_2;\epsilon+\omega)\Gb^{\text{A}}(\boldsymbol{x}_2,\boldsymbol{x}_1;\epsilon)F(\epsilon)\rangle\right.\non\\
  &\ \ \ \ \ \ \ \ +\left.\langle \Gb^{\text{R}}(\boldsymbol{x}_1,\boldsymbol{x}_2;\epsilon+\omega)\Gb^{\text{A}}(\boldsymbol{x}_2,\boldsymbol{x}_1;\epsilon)(F(\epsilon)-F(\epsilon+\omega))\rangle\right]\non\\
  =&-2i\,\Pi(\boldsymbol{x}_1,\boldsymbol{x}_2;\omega)
\end{align}
     %      \begin{align}
     %           &2\Pi(\boldsymbol{x}_1,\boldsymbol{x}_2;\omega)\\
              %               =&\frac{i}{2}\int\frac{d\epsilon}{2\pi}\left[\tr\langle G^{\text{R}}(\boldsymbol{x}_1,\boldsymbol{x}_2;\epsilon+\omega)G^{\text{R}}(\boldsymbol{x}_2,\boldsymbol{x}_1;\epsilon)F(\epsilon+\omega)-G^{\text{A}}(\boldsymbol{x}_1,\boldsymbol{x}_2;\epsilon+\omega)G^{\text{A}}(\boldsymbol{x}_2,\boldsymbol{x}_1;\epsilon)F(\epsilon)\rangle\right.\\
              %               &\ \ \ \ \ \ \ \ +\left.\tr\langle G^{\text{R}}(\boldsymbol{x}_1,\boldsymbol{x}_2;\epsilon+\omega)G^{\text{A}}(\boldsymbol{x}_2,\boldsymbol{x}_1;\epsilon)(F(\epsilon)-F(\epsilon+\omega))\rangle\right]
                                %         \end{align}

                                %         The factor of $2$ in the first line comes from the trace over the two valleys. 
$\Pi(\boldsymbol{x}_1,\boldsymbol{x}_2;\omega)$ is the disorder averaged polarization operator, note that the disorder average should be performed for both single particle Green's function and four-point correlation (i.e. the ladder diagrams), and its Fourier component is given by:
\begin{equation}
  \Pi(\boldsymbol{q},\omega)=\nu+\frac{\pi^2}{3}\nu^{\prime\prime}T^2+\nu\frac{i\omega}{Dq^2-i\omega},\ \ \ \ \ \text{for }q<l^{-1},\ \omega<\tau^{-1}
\end{equation}
Thus, in momentum $\boldsymbol{q}$ and frequency $\omega$ space, the quadratic term is given by:
\begin{equation}
  \boxed{-\frac{1}{2}U^2\tr\langle\left[\Gb\sigma_z\gamma^{\alpha}\frac{\Phi_{\alpha}}{2}\right]^2\rangle=iU^2\int\frac{d\omega}{2\pi}\int\frac{d^2q}{(2\pi)^2}\Phi_{\text{q}}(-\boldsymbol{q},-\omega)\Pi(\boldsymbol{q},\omega)\Phi_{\text{cl}}(\boldsymbol{q},\omega)}
\end{equation}
In the static limit, i.e. $\omega=0,\ve{q}\rightarrow 0$, we obtain the standard expression for $T_c$ of valley polarization as $1-U(\nu+\frac{\pi^2}{3}\nu^{\prime\prime}T_c^2)=0\rightarrow T_c=\sqrt{\frac{U\nu -1}{U\pi^2|\nu^{\prime\prime}|/3}}$.

\subsection{The cubic and the quartic term}

For completeness, we present the calculation for the cubic and the quartic terms.

The cubic term can be very similarly written down:
\begin{equation}
  \begin{split}
    &\frac{1}{3}U^3\tr\langle\left[\left(G_0^{-1}-V_{\text{dis}}\gamma^{\text{cl}}\right)^{-1}\frac{1}{2}\Phi_{\alpha}\sigma_z\gamma^{\alpha}\right]^3\rangle\\
    =&U^3\int dt_1dt_2dt_3\int d^2x_1d^2x_2d^2x_3\frac{1}{8}\Phi_{\text{q}}(\boldsymbol{x}_2,t_2)\Phi_{\text{cl}}(\boldsymbol{x}_3,t_3)\Phi_{\text{cl}}(\boldsymbol{x}_1,t_1)\\
    \times&\tr\langle G(\boldsymbol{x}_1,\boldsymbol{x}_2;t_1,t_2)\sigma_z\gamma^{\text{q}}G(\boldsymbol{x}_2,\boldsymbol{x}_3;t_2,t_3)G(\boldsymbol{x}_3,\boldsymbol{x}_1;t_3,t_1)\rangle
  \end{split}
\end{equation}
Notice that the last line should have been $\sim G\sigma_z\gamma^{\text{q}}G\sigma_z\gamma^{\text{cl}}G\sigma_z\gamma^{\text{cl}}$, which could be simplified.

The steps to proceed:
\begin{enumerate}
\item Put all the $\Phi$ fields at the same space-time point $(\boldsymbol{x}_2,t_2)$. With this approximation, we neglect the nonlocal effects. We choose the space-time coordinate of $\Phi_{\text{q}}$ as a reference point;
\item Expand the Keldysh Green's function as 
  \begin{equation}
    \begin{split}
      G^{\text{K}}(\boldsymbol{x}_1,\boldsymbol{x}_2;t_1,t_2)=&\int dt_3\int d^2x_3 \left[G^{\text{R}}(\boldsymbol{x}_1,\boldsymbol{x}_3;t_1,t_3)F(\boldsymbol{x}_3,\boldsymbol{x}_2;t_3,t_2)-F(\boldsymbol{x}_1,\boldsymbol{x}_3;t_1,t_3)G^{\text{A}}(\boldsymbol{x}_3,\boldsymbol{x}_2;t_3,t_2)\right]
    \end{split}
  \end{equation}
\item The cubic term would reduce to:
  \begin{equation}
    \begin{split}
      \sim&\frac{1}{8}U^3\int dt_2d^2x_2\Phi_{\text{q}}(\boldsymbol{x}_2,t_2)h^2_{\text{cl}}(\boldsymbol{x}_2,t_2)\int dt_1dt_3dt_4\int d^2x_1d^2x_3d^2x_4\\
      \times&\tr\langle\sigma_z\left[F(4,2)G^{\text{R}}(2,3)G^{\text{R}}(3,1)G^{\text{R}}(1,4)-F(2,4)G^{\text{A}}(4,3)G^{\text{A}}(3,1)G^{\text{A}}(1,2)\right]\rangle
    \end{split}
  \end{equation}
  Here, $(i,j)$ is short for $(\boldsymbol{x}_i,\boldsymbol{x}_j;t_i,t_j)$;
\item The second line may be evaluated in Fourier space, giving rise to $\sim i2\pi \int\frac{d^2k}{(2\pi)^2}\partial^2_{\epsilon_{\boldsymbol{k}}}\left[f^{(1)}(\boldsymbol{k})-f^{(2)}(\boldsymbol{k})\right]$;
\item The third order term is:
  \begin{equation}
    \boxed{\sim\frac{-i}{8}U^3\int dt_2d^2x_2\Phi_{\text{q}}(\boldsymbol{x}_2,t_2)\Phi^2_{\text{cl}}(\boldsymbol{x}_2,t_2)\int\frac{d^2k}{(2\pi)^2}\partial^2_{\epsilon_{\boldsymbol{k}}}\left[f^{(1)}(\boldsymbol{k})-f^{(2)}(\boldsymbol{k})\right]}
  \end{equation}
\end{enumerate}

One should notice that the cubic term vanishes in equilibrium. Thus, it involves the weak electric field and higher order of VPOP. Thus, the cubic term is neglected in the main text, when we discuss the VPOP physics close to or above the critical temperature, $T_c$.

The quartic term can be evaluated in the same way as the cubic term:
\begin{equation}
  \begin{split}
    &-\frac{1}{4}U^4\tr\langle\left[\Gb\frac{1}{2}\Phi_{\alpha}\sigma_z\gamma^{\alpha}\right]^4\rangle\\
    =&-U^4\int dt_1dt_2dt_3dt_4\int d^2x_1d^2x_2d^2x_3d^2x_4\frac{1}{16}\Phi_{\text{q}}(\boldsymbol{x}_2,t_2)\Phi_{\text{cl}}(\boldsymbol{x}_3,t_3)\Phi_{\text{cl}}(\boldsymbol{x}_4,t_4)\Phi_{\text{cl}}(\boldsymbol{x}_1,t_1)\\
    \times&\tr\langle G(\boldsymbol{x}_1,\boldsymbol{x}_2;t_1,t_2)\gamma^{\text{q}}G(\boldsymbol{x}_2,\boldsymbol{x}_3;t_2,t_3)G(\boldsymbol{x}_3,\boldsymbol{x}_4;t_3,t_4)G(\boldsymbol{x}_4,\boldsymbol{x}_1;t_4,t_1)\rangle
  \end{split}
\end{equation}
The next steps fully parallel the previous analysis of the cubic term:
\begin{enumerate}
\item Put all the $\Phi$ fields at the same space-time point $(\boldsymbol{x}_2,t_2)$. With this approximation, we neglect the nonlocal effects, which does not alter our main conclusion. We choose the space-time coordinate of $\Phi_{\text{q}}$ as a reference point;
\item Expand the Keldysh Green's function as, e.g.\
  \begin{equation}
    G^{\text{K}}(\boldsymbol{x}_1,\boldsymbol{x}_2;t_1,t_2)=\int dt_3\int d^2x_3 \left[G^{\text{R}}(\boldsymbol{x}_1,\boldsymbol{x}_3;t_1,t_3)F(\boldsymbol{x}_3,\boldsymbol{x}_2;t_3,t_2)-F(\boldsymbol{x}_1,\boldsymbol{x}_3;t_1,t_3)G^{\text{A}}(\boldsymbol{x}_3,\boldsymbol{x}_2;t_3,t_2)\right];
  \end{equation}
\item The quartic term would reduce to:
  \begin{equation}
    \begin{split}
      \sim&-\frac{1}{8}U^4\int dt_2d^2x_2\Phi_{\text{q}}(\boldsymbol{x}_2,t_2)\Phi_{\text{cl}}^3(\boldsymbol{x}_2,t_2)\int dt_1dt_3dt_4dt_5\int d^2x_1d^2x_3d^2x_4d^2x_5\,\\
      \times&\langle\left[F(5,2)G^{\text{R}}(2,3)G^{\text{R}}(3,4)G^{\text{R}}(4,1)G^{\text{R}}(1,5)-F(2,5)G^{\text{A}}(5,3)G^{\text{A}}(3,4)G^{\text{A}}(4,1)G^{\text{A}}(1,2)\right]\rangle
    \end{split}
  \end{equation}
  Here, $(i,j)$ is short for $(\boldsymbol{x}_i,\boldsymbol{x}_j;t_i,t_j)$;
\item The second line may be evaluated in Fourier space, giving rise to $\sim \frac{i}{3!} \int\frac{d^2k}{(2\pi)^2}\partial^3_{\epsilon_{\boldsymbol{k}}}\left[f^{(1)}(\boldsymbol{k})+f^{(2)}(\boldsymbol{k})\right]$;
\item The fourth order term would be:
  \begin{equation}
    \boxed{\begin{split}
        &\sim-\frac{i}{48}U^4\int \diff t \diff^2x\Phi_{\text{q}}(\boldsymbol{x},t)\Phi^3_{\text{cl}}(\boldsymbol{x},t)\int\frac{d^2k}{(2\pi)^2}\partial^3_{\epsilon_{\boldsymbol{k}}}\left[f^{(1)}(\boldsymbol{k})+f^{(2)}(\boldsymbol{k})\right]\\
        &=i\frac{1}{24}U^4\nu^{\prime\prime}\int \diff t\diff^2x\Phi_{\text{q}}(\boldsymbol{x},t)\Phi^3_{\text{cl}}(\boldsymbol{x},t)
      \end{split}}
  \end{equation}
\end{enumerate}

\subsection{Equation of Motion}

Combining the calculation above, one can obtain the action given at the beginning of this section. The equation of motion for the valley polarization order parameter $\Phi_{\text{cl}}$ could also be read out as:
\begin{equation}
  \Delta n_0 + \left[U\Pi(q,\omega)-1\right]\Phi_{\text{cl}}+\alpha_3\Phi^2_{\text{cl}}+\frac{1}{24}U^3\nu^{\prime\prime}\Phi^3_{\text{cl}}=0
\end{equation}

\section{Modeling of Twisted Bilayer Graphene (TBG)}
\label{Sec:ModelingofTBG}

This section presents the necessary technical details of our modeling of the twisted bilayer graphene. Our modeling is based on BM's continuous model in Ref. \cite{bistritzer2011moire} and its generalization to the situation with an arbitrary smooth lattice deformation in Ref. \cite{balents2019general}.

\subsection{The Model Hamiltonian for TBG under uniaxial strain}

As an example, we focus on the electronic states near the K-point of layer 1, which can be well described by the following Hamiltonian:
\begin{equation}
  H^{(1K)}(\boldsymbol{k})=\begin{bmatrix}
    h^{(+)}_{-\theta_w/2}(\boldsymbol{k}) & T_b & T_{tr} & T_{tl}\\
    T_b^{\dagger} & h^{(+)}_{\theta_w/2}(\boldsymbol{k}+\boldsymbol{q}^{\prime}_b) & 0 &0\\
    T_{tr}^{\dagger} & 0 & h^{(+)}_{\theta_w/2}(\boldsymbol{k}+\boldsymbol{q}^{\prime}_{tr}) & 0\\
    T_{tl}^{\dagger} & 0 & 0 & h^{(+)}_{\theta_w/2}(\boldsymbol{k}+\boldsymbol{q}^{\prime}_{tl})
  \end{bmatrix}
  \label{Eq:ModelHamiltonian}
\end{equation}
where $\theta_w$ is the twist angle. As argued in the maintext, we are neglecting the anisotropy in the Dirac Hamiltonian. Therefore, the diagonal terms are given by:
\begin{equation}
  h^{(+)}_{\theta}(\boldsymbol{k})=\begin{bmatrix}
    m & v_{\text{D}}ke^{-i(\theta_{\boldsymbol{k}}-\theta)}\\
    v_{\text{D}}ke^{-i(\theta_{\boldsymbol{k}}-\theta)} & -m
  \end{bmatrix}
\end{equation}
where $k=\left|\boldsymbol{k}\right|$ and $\theta_{\boldsymbol{k}}=\tan^{-1}\frac{k_y}{k_x}$ are the magnitude and the polar angle of momentum $\boldsymbol{k}$ measured from the K-point, respectively; the diagonal element, $m$, is the mass term induced by the alignment with the substrate.

The inter-layer coupling are described by the off-diagonal terms in Eq.~(\ref{Eq:ModelHamiltonian}), given by \cite{bistritzer2011moire,balents2019general}:
\begin{equation}
  T_b=w\begin{bmatrix}
    1 & 1\\
    1 & 1
  \end{bmatrix},\ \ T_{tr}=we^{-i\boldsymbol{\mathcal{G}}^{\prime}(2)\cdot\boldsymbol{d}+i\boldsymbol{\mathcal{G}}(2)\cdot \boldsymbol{\tau}}\begin{bmatrix}
    1 & e^{i\frac{2\pi}{3}}\\
    e^{-i\frac{2\pi}{3}} & 1
  \end{bmatrix},\ \ 
  T_{tl}=we^{-i\boldsymbol{\mathcal{G}}^{\prime}(3)\cdot\boldsymbol{d}+i\boldsymbol{\mathcal{G}}3)\cdot \boldsymbol{\tau}}\begin{bmatrix}
    1 & e^{-i\frac{4\pi}{3}}\\
    e^{i\frac{4\pi}{3}} & 1
  \end{bmatrix},
\end{equation}
where $w$ is the inter-layer coupling strength; $\boldsymbol{\mathcal{G}}^{(\prime)}(2,3)$ are the reciprocal lattice vector of the graphene layer 1(2); $\boldsymbol{d}$ and $\boldsymbol{\tau}$ are vectors defining the twist in real space: $\boldsymbol{R}^{\prime}=M(\theta_w)(\boldsymbol{R}-\boldsymbol{\tau})+\boldsymbol{d}$. Here, the effect of any strain field is also neglected. The strain field will introduce corrections in the reciprocal lattice vectors. Thus, the corrections to the inter-layer couplings are on the order of strain strength.

More importantly are the momenta $\boldsymbol{q}^{\prime}_i$ with $i=1,2,3$ (or $\text{b, tr, tl}$), which connect the K-point in layer 1 to the adjacent K-point in layer 2, Fig.~\ref{Fig:StrainAndBZ}(b). Mathematically, the momenta $\boldsymbol{q}^{\prime}_i$ are given by:
\begin{equation}
  \boldsymbol{q}^{\prime}_i=\boldsymbol{q}_i-\boldsymbol{\mathcal{E}}\cdot\boldsymbol{K}_i;\ \ \  \boldsymbol{q}_i=\left|\boldsymbol{K}\right|\theta_w\times\left\{(0,-1),\ (\frac{\sqrt{3}}{2},\frac{1}{2}),\ (-\frac{\sqrt{3}}{2},\frac{1}{2})\right\}
\end{equation}
where $\boldsymbol{\mathcal{E}}$ is the uniaxial strain tensor; $\boldsymbol{K}_i$ are the momenta of the three K-points of layer 1. Here, we consider only layer 1 is strained, without loss of generality.

For the convenience of analytical calculation, one may assume the inter-layer coupling is weak and solve for the wavefunction perturbatively for the states around the K-point, $\left|\boldsymbol{k}\right|\ll\left|\boldsymbol{q}_i\right|$. The wavefunction will be used to evaluate the impurity scattering amplitudes and the rates.

\subsection{Impurity Potential and Impurity Average}

In this section, we present the details about the impurity potential and impurity averaging process in our simplified model calculation. For the convenience of analytical calculation, we assumed short ranged impurities (for simplicity), whose functional form in real space is given by:
\begin{equation}
  V(\boldsymbol{r})=V\delta(\boldsymbol{r}-\boldsymbol{R}_{\text{imp}}).
\end{equation}
Therefore, for a scattering process of plane waves with momentum transfer of $\Delta\boldsymbol{k}$, the scattering amplitude is given by the Fourier transformation:
\begin{equation}
  V(\Delta\boldsymbol{k})=Ve^{i\Delta\boldsymbol{k}\cdot\boldsymbol{R}_{\text{imp}}}
\end{equation}
Here the phase factor is kept explicitly. It will be important below when doing the disorder averaging in the huge moir\'e unit cell.

Let's focus on the scattering between the two valleys of different layers, indicated by the red arrows in Fig.~\ref{Fig:TBLGScattering}. We start the analysis by assuming only the sublattice A of graphene layer 2 is disordered and evaluating the scattering rates. Then, one should do the same analysis for the disorder to be on the other sublattice site and the other graphene layer and do an algebraic average over the all the scattering rates.

\begin{figure}[tb]
  \centering
  \includegraphics[scale=0.35]{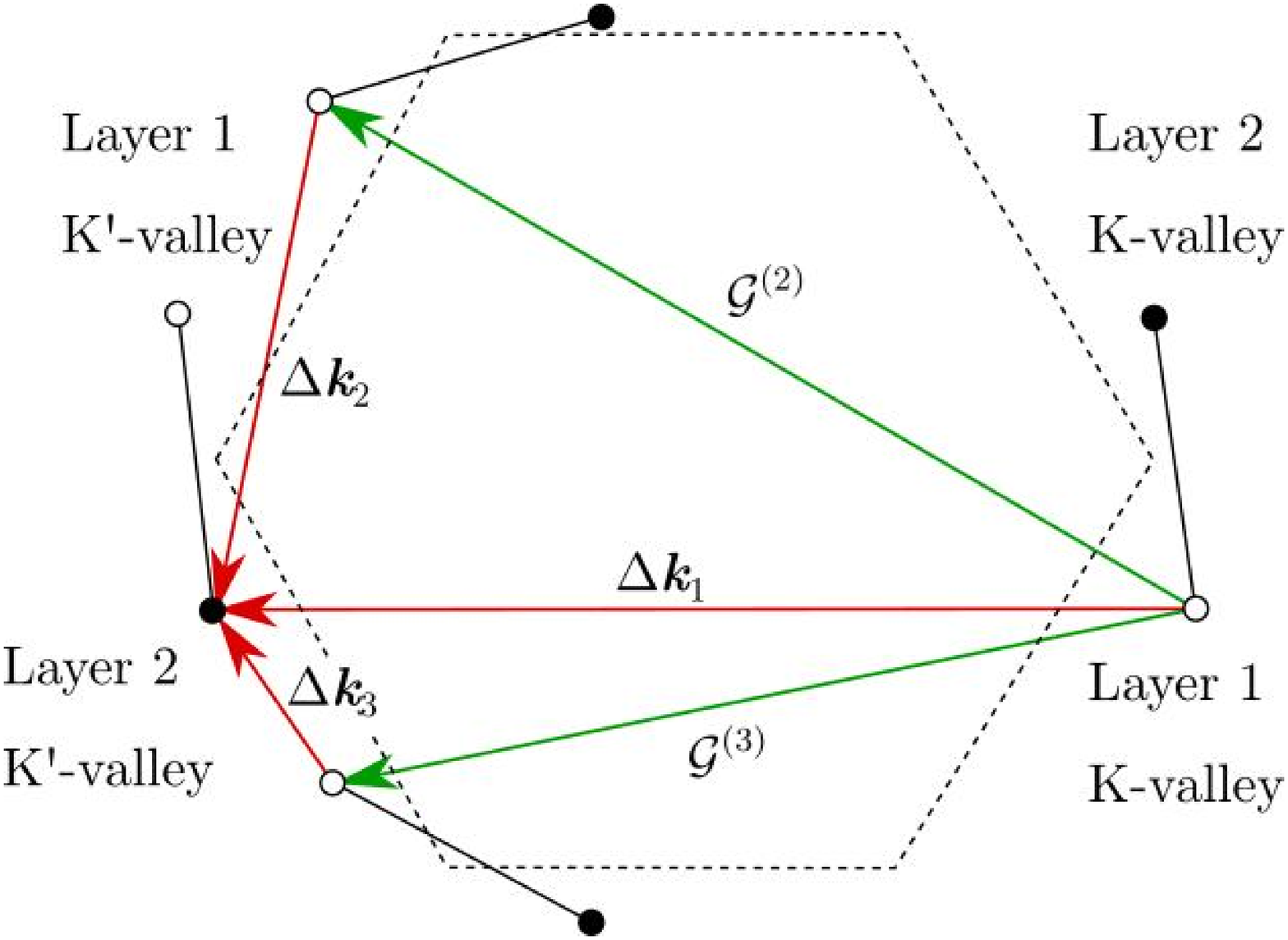}
  \caption{The possible momentum transfers $\Delta \boldsymbol{k}_i$ (red) upon impurity scattering between states of K-valley layer 1 and K$^{\prime}$ valley layer 2. The green arrows labels the reciprocal lattice vectors $\boldsymbol{\mathcal{G}}(2,3)$ of graphene layer 1.}
  \label{Fig:MomentumTransfer}
\end{figure}

Let's focus on the situation when only the sublattice A of graphene layer 2 is disordered. The scattering amplitude from a state $\boldsymbol{k}$ near the K-point of layer 1 to a state $\boldsymbol{k}^{\prime}$ near the K$^\prime$-point of layer 2 is given by:
\begin{equation}
  V^{(2K^{\prime})(1K)}_{\boldsymbol{k}^{\prime}\boldsymbol{k}}=\psi^{(2K^{\prime})\dagger}_{\boldsymbol{k}^{\prime}}\hat{V}^{(2A)}\psi^{(1K)}_{\boldsymbol{k}}
\end{equation}
where $\psi^{(1K)}_{\boldsymbol{k}}$ and $\psi^{(2K^{\prime})}_{\boldsymbol{k}^{\prime}}$ are the wavefunctions of the states near K-point of layer 1 and K$^{\prime}$-point of layer 2 correspondingly. The impurity matrix is given by an 8-by-8 matrix:
\begin{equation}
  \hat{V}^{(2A)}=V\left[
    \begin{array}{c | c}
      0_{2\times 2} & \begin{array}{c  c |  c  c |  c  c}
                        e^{i\Delta \boldsymbol{k}_1\cdot\boldsymbol{R}_{\text{imp}}} & 0 & e^{i\Delta \boldsymbol{k}_2\cdot\boldsymbol{R}_{\text{imp}}} & 0 & e^{i\Delta \boldsymbol{k}_3\cdot\boldsymbol{R}_{\text{imp}}} & 0\\
                        0 & 0 & 0 & 0 & 0 & 0
                      \end{array}\\ \hline
      0_{6\times2} & 0_{6\times 6}
    \end{array}\right]
  \label{Eq:ImpurityPotentialMatrix}
\end{equation}
The corresponding scattering rate is defined as:
\begin{equation}
  W^{(2K^{\prime})(1K)}_{\boldsymbol{k}^{\prime}\boldsymbol{k}}=2\pi \rho_{\text{imp}}\left|V^{(2K^{\prime})(1K)}_{\boldsymbol{k}^{\prime}\boldsymbol{k}}\right|^2
  \label{Eq:ScatteringRateFirstStep}
\end{equation}
where $\rho_{\text{imp}}$ is the impurity density.

Notice that the moir\'e lattice has a huge unit cell. Therefore, Eq.~(\ref{Eq:ScatteringRateFirstStep}) is not yet the scattering rate to be put in the Boltzmann equation. The scattering rate to be used in the Boltzamnn equation is obtained from Eq.~(\ref{Eq:ScatteringRateFirstStep}) by averaging over the impurity location $\boldsymbol{R}_{\text{imp}}$. The disorder averaging can be easily done by noticing that the phases in the impurity matrix, Eq.~(\ref{Eq:ImpurityPotentialMatrix}), are completely random relative to each other. Algebraic, this can be seen by factoring out the factor $e^{i\Delta \boldsymbol{k}_1\cdot\boldsymbol{R}_{\text{imp}}}$ in Eq.~(\ref{Eq:ImpurityPotentialMatrix}). The remaining phases involve the momentum difference. Notice that
\begin{equation}
  \Delta\boldsymbol{k}_{2}-\Delta\boldsymbol{k}_1=\boldsymbol{\mathcal{G}}(2),\ \ \Delta\boldsymbol{k}_{3}-\Delta\boldsymbol{k}_1=\boldsymbol{\mathcal{G}}(3)
\end{equation}
where $\boldsymbol{\mathcal{G}}(2/3)$ are the reciprocal lattice constant of graphene layer 1. Notice that the impurities are now assumed to be in graphene layer 2. The relative phases can be written as:
\begin{equation}
  \begin{split}
    e^{i\left(\Delta \boldsymbol{k}_{2/3}-\Delta \boldsymbol{k}_1\right)\cdot\boldsymbol{R}_{\text{imp}}}=&e^{i\boldsymbol{\mathcal{G}}(2/3)\cdot\boldsymbol{R}_{\text{imp}}}=e^{i\left[\boldsymbol{\mathcal{G}}(2/3)-\boldsymbol{\mathcal{G}}^{\prime}(2/3)\right]\cdot\boldsymbol{R}_{\text{imp}}}\\
    =&e^{i\boldsymbol{\mathcal{G}}^M(2/3)\cdot\boldsymbol{R}_{\text{imp}}}.
  \end{split}
\end{equation}
In the last equally of the first line, the reciprocal lattice vector of graphene layer 2, $\boldsymbol{\mathcal{G}}^{\prime}(2/3)$ is inserted. To go to the second line, one should notice that the difference of the reciprocal lattice vector of the two graphene layers defines the the reciprocal lattice vector of the moir\'e lattice, $\boldsymbol{\mathcal{G}}^M(2/3)$. At this point, due to the huge moir\'e unit cell, it's obvious that the relative phases in Eq.~(\ref{Eq:ImpurityPotentialMatrix}) are completely random upon impurity averaging. The disorder averaging of the scattering rates can be done by averaging the random phases:
\begin{equation}
  \langle W^{(2K^{\prime})(1K)}_{\boldsymbol{k}^{\prime}\boldsymbol{k}}\rangle_{\text{imp}}=2\pi \rho_{\text{imp}}\langle\left|V^{(2K^{\prime})(1K)}_{\boldsymbol{k}^{\prime}\boldsymbol{k}}\right|^2\rangle_{\boldsymbol{R}_{\text{imp}}}
\end{equation}
with $\boldsymbol{R}_{\text{imp}}$ being the location of the impurities, whose values correspond to the locations of the sublattice A of graphene layer 2. Algebraically, this disorder averaging process is the same as independently treating each nonzero element in Eq.~(\ref{Eq:ImpurityPotentialMatrix}) and calculating the scattering rates and then doing an averaging.

\subsection{Scattering Rates}

Following the procedure in the previous subsection, we were able to calculate the impurity scattering rate and extract the valley density difference under a DC current with numerical calculation. The result is summarized as the dotted line in Fig.~(\ref{Fig:result}).

Under certain limit, analytical expressions can be found to help understand the limiting factors of the valley density difference under a DC current. Below, we present the scattering rates under the limit of $v_{\text{D}}\left|
  \boldsymbol{k}^{(\prime)}\right|=v_{\text{D}}k_{\text{F}}\ll v_{\text{D}}\left|\boldsymbol{q}_i\right|\ll m$ and $\epsilon\ll\theta_w$ as well as the weak inter-layer coupling limit. The first condition, $v_{\text{D}}k_{\text{F}}\ll v_{\text{D}}\left|\boldsymbol{q}_i\right|\ll m$, states that the chemical potential is close to the bottom (top) of the conduction (valence) band so that the Fermi surfaces are approximately circular. The second condition, $\epsilon\ll\theta_w$, assumes weak strain strength so that the the rotational symmetry is weakly broken. The last simplification of weak inter-layer coupling manifests itself as the condition of:
\begin{equation}
  t=\frac{w}{v_{\text{D}}^2\boldsymbol{q}_i^2/2m}\ll 1.
  \label{Eq:WeakTunneling}
\end{equation}
Under the conditions stated above, we were able to find the leading order contribution to the inter-valley, inter-layer scattering rates (indicated as red arrows in Fig.~(\ref{Fig:TBLGScattering})) for electrons in the conduction band:
\begin{equation}
  \begin{split}
    W^{(2K^{\prime})(1K)}_{\boldsymbol{k}^{\prime}\boldsymbol{k}}\approx \frac{1}{\nu\tau^{\prime}}\frac{3}{4}t^2
    &\left\{1+\left[3\left(\epsilon_{xx}-\epsilon_{yy}\right)\frac{\epsilon}{\theta_w}\frac{k_{\text{F}}}{q}-3\epsilon_{xy}\frac{\epsilon}{\theta_w}\frac{v_{\text{D}}k_{\text{F}}}{m}\right]\cos\theta_{\boldsymbol{k}}+\left[-6\epsilon_{xy}\frac{\epsilon}{\theta_w}\frac{k_{\text{F}}}{q}-\frac{3}{2}\left(\epsilon_{xx}-\epsilon_{yy}\right)\frac{\epsilon}{\theta_w}\frac{v_{\text{D}}k_{\text{F}}}{m}\right]\sin\theta_{\boldsymbol{k}}\right.\\
    &\ \ \ \ \left.+\left[-3\left(\epsilon_{xx}-\epsilon_{yy}\right)\frac{\epsilon}{\theta_w}\frac{k_{\text{F}}}{q}-3\epsilon_{xy}\frac{\epsilon}{\theta_w}\frac{v_{\text{D}}k_{\text{F}}}{m}\right]\cos\theta_{\boldsymbol{k}^{\prime}}+\left[6\epsilon_{xy}\frac{\epsilon}{\theta_w}\frac{k_{\text{F}}}{q}-\frac{3}{2}\left(\epsilon_{xx}-\epsilon_{yy}\right)\frac{\epsilon}{\theta_w}\frac{v_{\text{D}}k_{\text{F}}}{m}\right]\sin\theta_{\boldsymbol{k}^{\prime}} \right\}
  \end{split}
  \label{Eq:InterVInterLScattering}
\end{equation}

% \end{widetext}

\end{document}